\documentclass[preprint,12pt]{elsarticle}

\usepackage{amssymb}
\usepackage[dvipsnames]{xcolor}
\usepackage{listings}
\usepackage{subcaption}
\usepackage{tikz}
\usepackage{float}
\usepackage{hyperref}
\definecolor{DarkRed}{rgb}{0.55,0,0}
\definecolor{DarkBlue}{rgb}{0,0,0.55}
\usepackage{algsimple}
\usepackage{lineno}

\journal{Computer-Aided Design}

\begin{document}

\begin{frontmatter}

\title{Tessellation and interactive visualization of four-dimensional spacetime geometries.}
\author{Philip Claude Caplan}
\affiliation{organization={Middlebury College, Department of Computer Science},
            addressline={75 Shannon Street}, 
            city={Middlebury},
            postcode={05753}, 
            state={VT},
            country={USA}}

\begin{abstract}
This paper addresses two problems needed to support four-dimensional ($3d + t$) spacetime numerical simulations.
The first contribution is a general algorithm for producing conforming spacetime meshes of moving geometries.
Here, the surface points of the geometry are embedded in a four-dimensional space as the geometry moves in time.
The geometry is first tessellated at prescribed time steps and then these tessellations are connected in the parameter space of each geometry entity to form tetrahedra.
In contrast to previous work, this approach allows the resolution of the geometry to be controlled at each time step.
The only restriction on the algorithm is the requirement that no topological changes to the geometry are made (i.e. the hierarchical relations between all geometry entities are maintained) as the geometry moves in time.
The validity of the final mesh topology is verified by ensuring the tetrahedralizations represent a closed 3-manifold.
For some analytic problems, the $4d$ volume of the tetrahedralization is also verified.
The second problem addressed in this paper is the design of a system to interactively visualize four-dimensional meshes, including tetrahedra (embedded in $4d$) and pentatopes.
Algorithms that either include or exclude a geometry shader are described, and the efficiency of each approach is then compared.
Overall, the results suggest that visualizing tetrahedra (either those bounding the domain, or extracted from a pentatopal mesh) using a geometry shader achieves the highest frame rate, in the range of $20-30$ frames per second for meshes with about $50$ million tetrahedra.
\end{abstract}


\end{frontmatter}

\section{Introduction}
\label{sec:introduction}

\begin{figure}[H]
  \centering
  \begin{subfigure}{0.575\textwidth}
    \includegraphics[width=\textwidth]{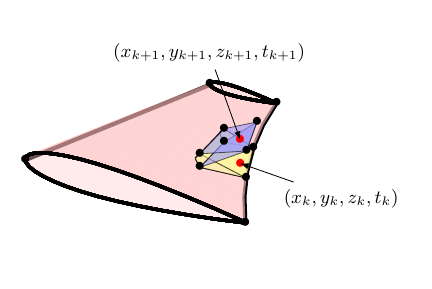}
  \end{subfigure}
  \begin{subfigure}{0.4\textwidth}
    \includegraphics[width=\textwidth]{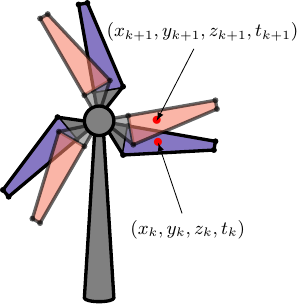}
  \end{subfigure}
  \caption{Examples of $3d$ geometries moving in spacetime ($4d$).
  The wind turbine geometry was adapted from a post on the \LaTeX\ Stack Exchange~\cite{Wibrow_2014}.}
  \label{fig:intro}
\end{figure}

Many physical phenomena exhibit time-dependent features that are challenging to predict when the domain of interest is bounded by complex geometries.
Some examples involve predicting the flow physics of a rotating wind turbine (Fig.~\ref{fig:intro} right) or over an aircraft wing when flaps are deployed (Fig.~\ref{fig:intro} left).
Numerical simulations offer time- and cost-efficient methods for predicting these flows and, within the realm of numerical simulations, \textit{spacetime} numerical simulations have gained recent attention due to their potential to reduce the overall cost of the computation~\cite{Yano_2012_PhD}.

Spacetime numerical simulations consist of embedding a three-dimensional ($3d$) geometry into four-dimensional ($4d$) space, where the fourth dimension is time ($t$ in Fig.~\ref{fig:intro}).
Finite element simulations can then be performed in this coupled space-time domain\cite{Yano_2012_PhD, Argyris_1969_Space_Time_FE, Behr_2008_Simplex_spacetime_meshes_FEM,Yano_2014_Spacetime_Adaptation,Jayasinghe_2018_PhD, Caplan_2019_PhD}.
Visualizing the mesh and resulting solution fields associated with these $4d$ simulations is necessary to be able to both debug meshing techniques and investigate features in the solution.
Four-dimensional visualization techniques can be categorized as either projection-based or intersection-based, both of which use a $4d$ camera to transform the $4d$ objects into a $3d$ viewing space which can then be rendered with standard $3d$ techniques~\cite{Cavallo_2021}.
In projection-based methods, points are transformed to the $3d$ viewing space using a projection matrix, for example, defined by a perspective projection transformation.
This approach is commonly used to visualize a tesseract in which one of the bounding cubes of the tesseract appears smaller~\cite{Cavallo_2021} since it is further from the observation point.
Intersection-based methods (or cross-sections~\cite{Cavallo_2021}) involve slicing the four-dimensional tetrahedra or pentatopes with a four-dimensional hyperplane.
Any mesh edges that cross the hyperplane can be intersected with the plane, thus providing three-dimensional coordinates in the basis of the hyperplane or, equivalently, in the frame of reference of the $4d$ camera.
This also requires tessellating the portion of the cell (tetrahedron or pentatope) that is intersected by the hyperplane.
Previous work involves computing these intersections as a pre-processing step, and then rendering the resulting polyhedra \cite{Caplan_2019_PhD} or tetrahedra \cite{BeldaFerrin_2020}.
In the current work, the intersection-based method is preferred over the projection-based approach since it can be used to exactly visualize the geometry at specific times.
This concept has been explored by Zhang~\cite{Zhang_2020} in the development of a $4d$ physics game engine.
Zhang uses a compute shader to calculate the intersection regions interactively and render simple shapes defined by relatively few tetrahedral elements.
The goal, in this work, is to visualize meshes with tens of millions of elements while ensuring interactivity is maintained.
Whenever the hyperplane definition is changed (e.g. by user input), the primitives that need to be passed to the rendering system change.
While they can be rebuffered to the GPU, this can slow down the rendering of each frame which impacts interactivity.
Thus, the primary goal is to develop a system for interactively visualizing four-dimensional meshes, where interactivity is defined (in this work) as achieving a frame rate of $20-30$ frames per second for relatively large meshes - about $50$ million (M) elements.
But, in order to study such techniques, we need meshes to visualize.
Thus, a method for generating fully unstructured $4d$ tetrahedral meshes of moving geometries has also been developed.

Work in unstructured $4d$ meshing has primarily focused on mesh adaptation~\cite{Caplan_2019_PhD, Gruau_2005, Tremblay_2007_Anisotropic_mesh_adaptation_for_time-continuous_spacetime_FEM_incompressible_NS} whereby an initial pentatope mesh is adapted via point insertion, edge collapse, edge flips and vertex smoothing, all within the cavity operator framework introduced by Coupez and Loseille~\cite{Coupez_2000_Parallel_meshing_and_remeshing, Coupez_2000_Generation_de_maillage_et_adaptation_par_optimisation_locale, Loseille_2013_Cavity_based_operators_for_mesh_adaptation, Loseille_2017_Unique_cavity-based_operator_and_hierarchical_domain_partitioning_for_fast_parallel_generation_of_anisotropic_meshes}.
The work of Caplan~\cite{Caplan_2019_PhD, Caplan_2020_CAD} assumes an initial $4d$ mesh can be obtained, using the Coxeter-Kuhn-Freudenthal triangulation~\cite{Kuhn_1968} as a starting point, which restricts the simulations to a tesseract domain.
An important question remaining from the aforementioned work is: \emph{how can we generate an initial mesh for complex, moving geometries?}.

Existing work in $4d$ mesh \emph{generation} includes extrusion-based~\cite{Behr_2008_Simplex_spacetime_meshes_FEM, Karabelas_2015}, elasticity-based~\cite{Hilger_2018, vonDanwitz_2021} or advancing front~\cite{Ungor_2000_Tent_pitcher, Erickson_2005_Building_space-time_meshes_over_arbitrary_spatial_domains, Gopalakrishnan_2016_Mapped_tent_pitching_schemes_for_hyperbolic_systems, Mont_2011_Adaptive_Unstructured_FourD_Spacetime_DG, Thite_2007_Adaptive_Spacetime_Meshing_DG} approaches.
For example, Behr~\cite{Behr_2008_Simplex_spacetime_meshes_FEM} creates prisms by extruding an initial spatial mesh - prisms are then subdivided into tetrahedra by inserting vertices along the temporal direction.
In elasticity-based approaches, an initial $4d$ mesh can be modeled as an elastic solid - vertex coordinates can then be moved to conform to the moving domain boundaries (with topology changes) while keeping the mesh topology fixed~\cite{vonDanwitz_2021}.
The Tent Pitcher algorithm of \"Ung\"or and Sheffer~\cite{Ungor_2000_Tent_pitcher} (extended to $4d$ by Mont~\cite{Mont_2011_Adaptive_Unstructured_FourD_Spacetime_DG}) is similar to advancing front methods.
This method starts from an initial spatial mesh and inserts points in the temporal direction to satisfy constraints on the dihedral angles of the internal mesh faces with respect to the spatial domain.
Other techniques locally modify the mesh topology to accommodate the geometry motion~\cite{Wang_2015}.
The aforementioned methods have focused primarily on the generation of pentatope meshes but are restricted in the way complex geometries with large movements are handled.

The current problem definition consists of tetrahedralizing the volume traced by a moving geometry in $4d$.
The most recent work in this area is that of Anderson et al.~\cite{Anderson_2023}, in which spacetime meshes of moving geometries are created by dividing the temporal interval of the domain into discrete slabs.
Within a single slab, the coordinates of each (surface) mesh vertex from the previous time step are advanced according to the local geometry velocity and, finally, projected to the moving geometry at the next time step.
Surface mesh triangles at the initial and terminating temporal planes are then connected to form prisms, maintaining the number of vertices and surface mesh topology from one time step to the next.
The prisms are subsequently subdivided to form tetrahedra in $4d$, whereby each prism is subdivided into 14 tetrahedra (though the authors note a conformal subdivision into 10 tetrahedra is possible), thus introducing several vertices in between the initial and terminating planes of each time slab.
Furthermore, Ko and Sakkalis remark upon the difficulty of projecting to a CAD geometry in the most general settings, since \textit{there is no perfect algorithm to solve the problem of orthogonal projection that satisfies high accuracy, robustness and efficient computation time simultaneously}~\cite{Ko_2014}.
Ko and Sakkalis also mention that care needs to be taken in degenerate cases near singularities.

Here, the goal is to develop a general algorithm for creating fully unstructured spacetime tetrahedral meshes of complex, moving geometries that can finally be used to evaluate the developed interactive visualization system.
The proposed method minimizes the addition of vertices in between time steps and avoids projecting to the CAD geometry.
Furthermore, the resolution of the geometry (and resulting topology of the surface meshes) is also allowed to change over time, thus demonstrating a fully unstructured spacetime mesh generation algorithm for $3d+t$ ($4d$) geometries.
It should be noted that the focus here is on producing conforming \emph{tessellations} of the spacetime geometries so they can be visualized - improving the quality of these meshes for numerical simulations is left for future work.
The correctness of the algorithm is verified by measuring the volume of the resulting tetrahedra for simple analytic problems.
The tessellation algorithm is also demonstrated on more complex geometries that include sharp features, (e.g. at the tip or trailing edge of a wing) and singularities (e.g. at the poles of a sphere).
Finally, these meshes enable the evaluation of several techniques for interactively visualizing a moving geometry, whereby the meshes (either tetrahedral or pentatopal) are intersected with a user-defined hyperplane.
Solutions with or without the use of geometry shaders~\cite{Wolff_2018} are presented, and the interactivity is investigated by measuring the resulting frame rate.

\section{Tessellating moving geometries.}
\label{sec:tessellate}

The proposed method for tessellating the volume traced by a moving $3d$ geometry is an extension of classical algorithms for tessellating $3d$ surfaces consisting of several patches.
These algorithms typically start by discretizing the CAD Edges, and then fix these discretizations when tessellating each surface patch (Face).
Each surface patch references the unique vertices in the discretization of the CAD Edges, as well as those placed on the CAD Nodes, thus providing a conformal tessellation of the entire geometry.
Please note that, in the current work, CAD Nodes, Edges and Faces are distinguished from mesh nodes (usually, vertices), edges, and faces (usually, triangles) by the capitalization of the first letter of each entity.

\begin{figure}
  \centering
  \includegraphics[width=\textwidth, trim={0 1cm 0 1cm}]{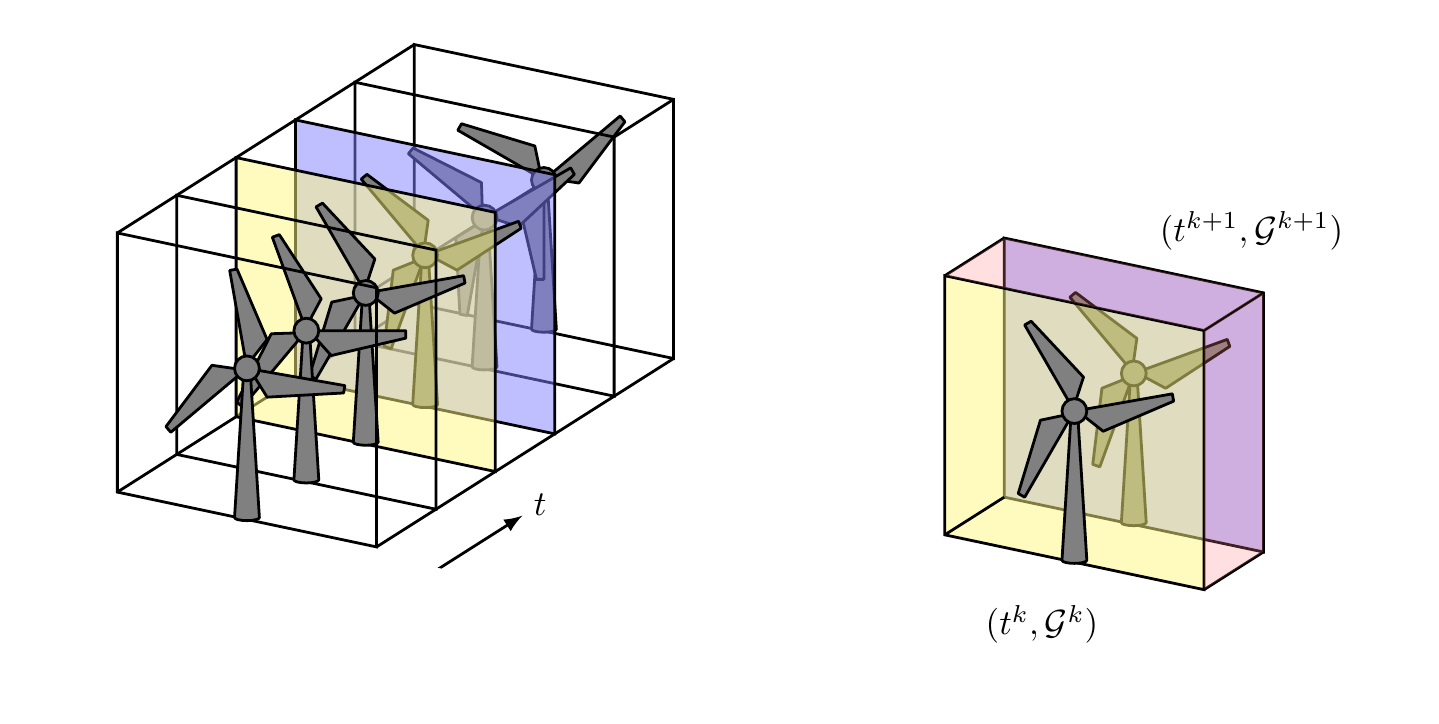}
  \caption{Illustration of how the time-dependent geometry is tessellated at discrete time steps (left) and how a time slab is defined between time steps (right).}
  \label{fig:slabs}
\end{figure}

First, the temporal axis is discretized into $n$ discrete slabs (see Fig.~\ref{fig:slabs}) and it is assumed that the time-dependent geometry can be determined at some time $t$.
It is also assumed that a $3d$ tessellation of the geometry can be obtained at time steps $t^{k}$ and $t^{k+1}$, i.e. a tessellation at all $n+1$ time steps.
The goal is to connect the tessellations at time steps $t^k$ and $t^{k+1}$ which should be understood to lie on the temporal boundaries of the current time slab.
The inputs are thus (1) a tessellation $\mathcal{T}^k(\mathcal{G}(t^k))$ for the geometry $\mathcal{G}$ evaluated at time $t$, and (2) another tessellation $\mathcal{T}^{k+1}(\mathcal{G}(t^{k+1}))$ for the geometry evaluated at time $t^{k+1}$.
Note that on the next time slab, $\mathcal{T}^{k+1}(\mathcal{G}(t^{k+1}))$ becomes the first input, thus ensuring conformity between neighboring time slabs.
For brevity, the notation $(\mathcal{G})$ will be dropped when denoting the tessellations - tessellations at the boundaries of each slab will simply be written as $\mathcal{T}^k$ and $\mathcal{T}^{k+1}$.

Each tessellation $\mathcal{T}^k$ is composed of the tessellations of the underlying CAD Faces, Edges and Nodes, which are denoted as $\mathcal{F}^k$, $\mathcal{E}^k$ and $\mathcal{N}^k$ at time step $t^k$.
At the next time step $t^{k+1}$, these are denoted as $\mathcal{F}^{k+1}$, $\mathcal{E}^{k+1}$ and $\mathcal{N}^{k+1}$ which form the tessellation $\mathcal{T}^{k+1}$.
Assuming a simplicial tessellation, $\mathcal{F}^k$ is composed of triangles, $\mathcal{E}^k$ is composed of edges and $\mathcal{N}^k$ is a collection of vertices.
Each of these is further subscripted with the unique index of each Node, Edge or Face in the geometry.
For example, $\mathcal{F}_i^k$ might be one of the surface patches of the wind turbine blade at time $t^k$ and $\mathcal{F}_j^k$ might be one of the surface patches on the cylindrical mast of the wind turbine at time $t^k$.
Note that $\mathcal{N}^k_i$ is a single vertex.

\begin{figure}
  \centering
  \includegraphics[width=\textwidth, trim={1cm 0 2cm 0}]{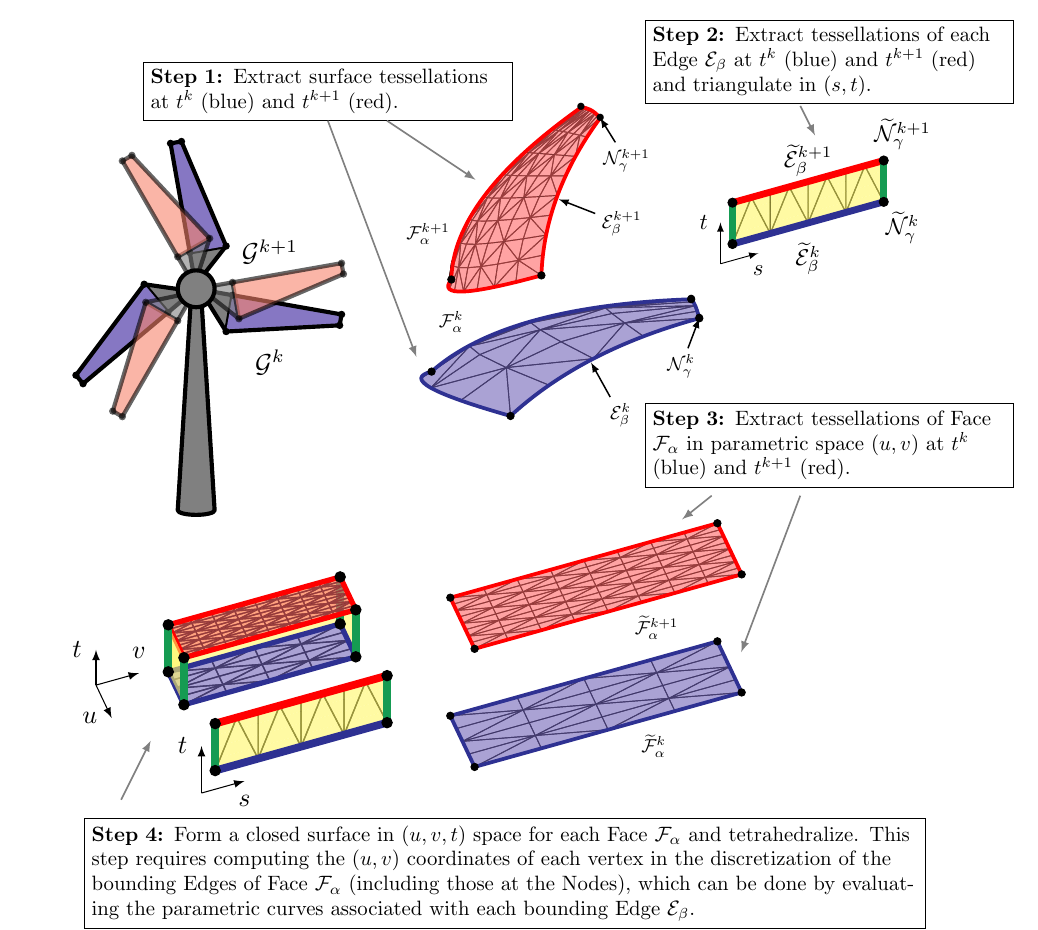}
  \caption{Description of the algorithm for tessellating a moving geometry within a time interval $[t^k, t^{k+1}]$.
    When a CAD Face ($\mathcal{F}$), Edge ($\mathcal{E}$) or Node $(\mathcal{N}$) is overscored with a tilde (\ $\widetilde{}$\ ), it should be understood that the entity is in some parametric space, whether it be that of an Edge ($s$) or Face $(u, v)$.
    A Node $\mathcal{N}$ can be mapped to the parametric spaces of either an Edge $\mathcal{E}$ or Face $\mathcal{F}$.
    The subscripts ($\alpha,\ \beta,\ \gamma$) denote the unique identifiers assigned to each entity.
  }
  \label{fig:tessellation}
\end{figure}

The approach for tessellating a moving geometry (illustrated in Fig.~\ref{fig:tessellation}) consists of connecting each tessellation from time $t^k$ to its sibling tessellation at time $t^{k+1}$.
For example, connecting $\mathcal{N}^k_\gamma$ to $\mathcal{N}^{k+1}_\gamma$ traces a $4d$ curve, which can be discretized with a single edge.
Connecting $\mathcal{E}^k_\beta$ to $\mathcal{E}^k_\beta$ traces $4d$ surface which can be triangulated, and connecting $\mathcal{F}^k_\alpha$ to $\mathcal{F}^{k+1}_\alpha$ traces a $4d$ volume to be tetrahedralized.
Note that $\alpha$, $\beta$ and $\gamma$ denote the unique integer identifiers assigned to each geometry entity.
The method to tessellate the moving Nodes (resulting in a $4d$ curve), Edges (resulting in a $4d$ surface) and Faces (resulting in a $4d$ volume) are described in the following sections.

\subsection{Tessellating Nodes (curves) and Edges (surfaces)}
\label{sec:tess-edge}
The discretization of each Node within a single spacetime slab is trivial since it simply consists of a single edge connecting the two vertices placed on $\mathcal{N}^k_\gamma$ and $\mathcal{N}^{k+1}_\gamma$.
For each CAD Edge $\mathcal{E}_\beta$, the discretizations at times $t^k$ and $t^{k+1}$ are first extracted - these are the set of edges on Edge $\mathcal{E}_\beta$ at time steps $t^k$ and $t^{k+1}$.
The goal is then to triangulate the $4d$ surface traced by $\mathcal{E}_\beta$ as it moves between time steps.
The geometry engine used here (\texttt{EGADS})~\cite{Haimes_2013_ESP, Haimes_2018_EGADSlite_A_lightweight_geometry_kernel_for_HPC} provides the parametric coordinate of each vertex in the discretization of $\mathcal{E}_\beta$.
This is a single coordinate since $\mathcal{E}_\beta$ is a 1-manifold - this coordinate is typically represented as $t$ but is denoted as $s$ to avoid confusion with time in this article.
Combined with the edges connecting the Nodes bounding $\mathcal{E}_\beta$, the next step consists of extracting the set of edges defining a closed curve in a two-dimensional $(s, t)$ space.
The closed region bounded by this curve is then triangulated (Step 2 in Fig.~\ref{fig:tessellation}) - here, \texttt{Triangle}~\cite{Shewchuk_1996} is used, but note that this domain is actually rectangular, so simpler algorithms could be implemented.
This process is repeated for each Edge in order to produce triangulations of each Edge of the geometry.

\subsection{Tessellating Faces (volumes)}
\label{sec:tess-face}
Next, each CAD Face is treated in a similar fashion, whereby the goal is now to tetrahedralize the volume traced by each Face $\mathcal{F}_\alpha$ as it moves in $4d$.
Again, the geometry engine used here provides the parametric coordinates $(u, v)$ of each interior vertex in the disretizations of $\mathcal{F}^k_\alpha$ and $\mathcal{F}^{k+1}_\alpha$.
The boundary vertices in the tessellation of $\mathcal{F}_\alpha$ lie on the Edges bounding the Face and, therefore, only contain a single $s$ parameter coordinate.
Geometry kernels typically provide a function to compute the $(u, v)$ coordinates of a vertex on a Face, given some $s$ value along the Edge (i.e. evaluating the parametric curves associated with the Edge in the Face).
With \texttt{EGADS}, this can be achieved with the \texttt{getEdgeUV} function.

The triangulations of the Faces (now defined in parametric $(u, v)$ space) are now embedded into a $3d$ $(u, v, t)$ space by placing the vertices of $\mathcal{F}^k_\alpha$ at $t^k$ and those of $\mathcal{F}^{k+1}_\alpha$ at $t^{k+1}$ (Step 3 in Fig.~\ref{fig:tessellation}).
To define a closed surface, it is necessary to extract the triangulations of the CAD Edges that bound the Face (computed using the algorithm described in Section~\ref{sec:tess-edge}).
This closed surface now encloses the volume traced by the face as it moves from time $t^k$ to $t^{k+1}$, which is tetrahedralized using \texttt{TetGen}~\cite{Si_TetGen_2015} (Step 4 in Fig.~\ref{fig:tessellation}).

Note that it is very possible \texttt{TetGen} adds Steiner vertices in the \emph{interior} of the $(u, v, t)$ volume.
Each of these Steiner vertices lies in between the two time steps, for which an analytic description of the geometry is not available.
The first two coordinates define the $(u, v)$ coordinates along Face $\mathcal{F}_\alpha$, so $\mathcal{F}^{k}_\alpha$ and $\mathcal{F}^{k+1}_\alpha$ are both evaluated, and the weighted average of the results (weighted by $t$) is used to determine the $3d$ coordinates of this Steiner vertex, which are then embedded in $4d$.

\subsection{Ensuring a closed 3-manifold mesh.}
\label{sec:tess-closed}
Taking a step back to $3d$ for a moment - assume only manifold geometries are considered, meaning each CAD Edge is incident to exactly two CAD Faces.
When creating a volume mesh for a manifold $3d$ domain, the bounding surface mesh typically needs to be \textit{closed} - that is, each mesh edge in the surface mesh needs to be adjacent to exactly two boundary triangles.

The same idea extends to the $4d$ setting: the boundary of the $4d$ domain needs to consist of a tetrahedralization in which each \emph{triangle} is adjacent to exactly two tetrahedra.
Thus, the indices of the mesh vertices need to be carefully synchronized when the tetrahedralization of a time slab is appended to the final mesh.
Furthermore, two additional tetrahedralizations are required at the extreme boundaries of the domain along the temporal axis, i.e. at the initial and final times of the entire spacetime domain.
Once these two tetrahedralizations are obtained, they are appended to the final mesh and the 3-manifold property (each triangle is adjacent to exactly two tetrahedra) is checked to ensure the meshes are valid.

It should be noted, however, that it may not always be possible to recover the segments (edges) during the tetrahedralizations in Section~\ref{sec:tess-face}.
In particular, Steiner vertices may be left on boundary mesh edges, specifically if the parameter space triangulations contain very poor-quality triangles which are considered intersecting and thus, not recoverable by the constrained Delaunay tetrahedralization algorithm.
When this is detected, the original surface mesh is recovered (necessary to ensure the final mesh is valid) by simply deleting the Steiner vertex, hence collapsing the unwanted segment along with any attached tetrahedra.

\section{Efficiently visualizing tetrahedra and pentatopes in $4d$.}
\label{sec:visualization}

A $4d$ mesh can be visualized using either a projection-based approach or an intersection-based approach~\cite{Cavallo_2021}.
Here, the intersection-based approach is preferred since it directly lends to visualizing the geometry.
In this approach, a four-dimensional hyperplane $\mathcal{H}$ is defined by a normal direction $\vec{n}$ and a point on the plane $\vec{c}$.
Any mesh entities that contain vertices on either side of $\mathcal{H}$ define an intersection region that needs to be visualized.
As the definition of $\mathcal{H}$ changes (e.g. interactively by a user), the number of intersected mesh entities changes.
Thus, the main challenge consists of designing an efficient system that can compute these intersection regions and render them interactively.
Here, ``interactive" is defined such that the resulting frame rates should be at least 20 FPS for reasonably large meshes (about 50M elements).

One method for computing intersection regions within a rendering pipeline involves the use of a geometry shader to generate the rendering primitives defining the intersection.
In this approach, mesh entities (vertices, connectivities) can be written to texture units~\cite{Feuillet_2021} and a single invocation of the geometry shader can be used to process a single mesh entity (triangle, tetrahedron, pentatope).
Within the shader, the hyperplane $\mathcal{H}$ can then be used to compute whether the mesh element intersects $\mathcal{H}$ and then output the necessary primitives (either a line primitive for a mesh triangle or triangle primitives for a tetrahedron or pentatope).
As noted by Maunoury~\cite{Maunoury_2022}, the injection of a geometry shader into the rendering pipeline (even as a pass-through shader) can significantly reduce the resulting frame rate.
As a result, solutions that either include or exclude the use of a geometry shader are explored.
Both of these solutions rely on the same mechanics for computing intersection primitives, which are described in Section~\ref{sec:vis-intersection} below.

Note that the focus is on intersecting tetrahedra with a hyperplane $\mathcal{H}$ since pentatopes can be visualized by simply passing the five bounding $4d$ tetrahedra of a single pentatope to the same tetrahedron-hyperplane intersection algorithm.

\subsection{Computing intersection primitives.}
\label{sec:vis-intersection}
In this work, intersection primitives (also referred to as rendering primitives) are defined as the simplicial decomposition of the  intersection region between a mesh entity and the $4d$ hyperplane which is to be rasterized (e.g. with \texttt{OpenGL}).
Let us focus on the case in which the mesh entity is a tetrahedron since this defines the moving geometry of the previous section.
Given a tetrahedron $\kappa$ and a hyperplane $\mathcal{H}$, the task is to produce triangles (if any) that represent the intersection $\kappa \cap \mathcal{H}$.
Similar to a tetrahedron-plane intersection in $3d$, a tetrahedron-hyperplane intersection in $4d$ results in either (1) an empty intersection, (2) a triangle or (3) a quadrilateral intersection primitive.

\begin{figure}
  \centering
  \includegraphics[width=1.03\textwidth, trim={1.2cm 0 0 0}]{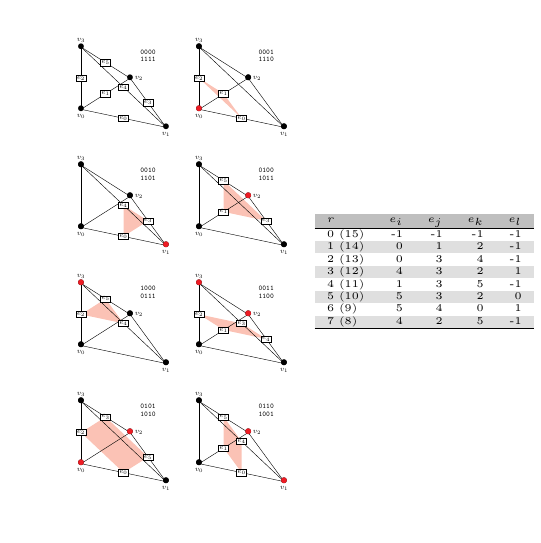}
  \caption{Description of the hyperplane-tetrahedron intersection algorithm.
  Intersections are listed in the same order as the Marching Tetrahedra algorithm~\cite{Bourke_1994}.
  Only 8 rows of the look-up table are shown due to symmetry (the symmetric values of $r$ are labelled in parentheses).
  The intersection region (to be decomposed into rendering primitives) are shown in red.
  For each case, red vertices are on one side of the hyperplane $\mathcal{H}$ whereas the black vertices are on the other side.
  }
  \label{fig:vis}
\end{figure}

Determining the intersection region can be reduced to determining which of the six edges of $\kappa$ intersect $\mathcal{H}$.
Either 0, 3 or 4 edges will be intersected for any given tetrahedron - the cases in which $\mathcal{H}$ exactly touches some of the tetrahedron vertices are treated in the same way (some of the points of the resulting rendering primitive will simply collapse onto each other).
A naive implementation would consist of using a series of conditional statements to determine which edges are intersected and how that produces a rendering primitive.
However, implementing these conditional statements within a shader (running on a GPU) can cause control divergence between the individual threads processing different tetrahedra, which can significantly impact performance~\cite{Halladay_2019}.
When the threads are grouped in such a way that the branches taken by each thread are the same, then performance will not necessarily be impacted.
While the likelihood that each thread group follows a similar branching pattern can be improved (e.g. by performing a spatial ordering of the input cells~\cite{Sander_2007, Han_2017}), there is still no guarantee that thread execution will not diverge from SIMD execution.

As a result, the algorithm developed here is designed to minimize the possibility of control divergence in the tetrahedron-hyperplane intersection routine.
The developed method is inspired by the Marching Tetrahedra algorithm~\cite{Bourke_1994} in which all possible tetrahedron-plane intersections are encoded as a four-bit number.
Look-up tables are then defined for the 16 possible cases that directly provide information as to how the intersection points should be calculated.

Fig.~\ref{fig:vis} describes the geometry and associated look-up table entries for each intersection, using the same order as the original Marching Tetrahedra algorithm~\cite{Bourke_1994}.
First, the \emph{side} of each of the four vertices of the tetrahedron is determined and is assigned to \texttt{0} if the vertex is on the positive side of the plane (red, in the direction of $\vec{n}$) and \texttt{1} otherwise (black).
The sides of all four vertices can be interpreted as a 4-bit number for the given tetrahedron, which represents which of the 16 cases needs to be addressed.
A $16 \times 4$ look-up table is then used to determine which edges are intersected for each case - the second dimension of the table is 4 because there are at most 4 edges intersected.
When the result corresponds to a triangle, the fourth entry is set in such a way that the quadrilateral still represents a triangle, thus avoiding a conditional to treat triangles or quadrilaterals (set to $-1$ in the table of Fig.~\ref{fig:vis}).
The only conditional appears when the result indicates an empty intersection (the tetrahedron lies entirely on one of the plane sides).
This is acceptable when tetrahedra are spatially close to each other since they are more likely to follow the same branching sequence (either no intersection or \emph{some} intersection).

Another table is then used to look up the tetrahedron vertices of each intersected edge (not shown in Fig.~\ref{fig:vis}), as well as a table to determine if the opposite edge of an intersection point is ``visible" (0 or 1).
The visibility value weights the altitude of the intersection point within a triangle rendering primitive.
The latter is used in the solid wireframe model~\cite{Brentzen_2006} to render the edges of the intersection primitive (the four edges of a quadrilateral or the three edges of a triangle).

The resulting intersection points are then $4d$ points that lie on $\mathcal{H}$.
Depending on the definition of the normal vector of $\mathcal{H}$, the $3d$ coordinates in the basis of $\mathcal{H}$ then form the $3d$ coordinates that are to be visualized.
These are transformed to clip space by the standard model-view-projection matrix, thus connecting the visualizer to user input such as zooming in/out, translating or rotating the meshes.

\subsection{Rendering $4d$ tetrahedra with a geometry shader.}
\label{sec:vis-gs}
A tetrahedral mesh embedded in $4d$ (e.g. produced by the algorithm described in Section~\ref{sec:tessellate}) can be visualized using the intersection algorithm described in the previous section.
First, the tetrahedron indices and vertex coordinates are buffered to the GPU so this data can be retrieved within the shaders using texture fetches.
The rendering pipeline is then invoked using \texttt{glDrawArrays} with the \emph{count} as the number of tetrahedra to render.
The vertex shader simply passes the invocation ID to the geometry shader.
The geometry shader then uses the invocation ID (the tetrahedron index) to compute the intersection region (triangles) using Alg.~\ref{alg:vis-geometry-shader}.
The output intersection primitives are then rasterized, leaving the fragment shader to evaluate the Phong reflection model with some specified diffuse reflection coefficient.
In the renderings presented in later sections, this diffuse reflection coefficient will often be a color associated with the unique identifier assigned to a geometry entity, which is used to distinguish between different surface patches.

\begin{algsimple}[H]
  \scriptsize
  \begin{algcode}[processTetrahedron]
    \zi \Inputs tetrahedron index ($\texttt{int}$), hyperplane definition: $\vec{n}$ (\texttt{vec4}), $\vec{c}$ (\texttt{vec4})
    \zi \Outputs intersection primitives (triangles)

    \li Retrieve vertex indices (\texttt{ivec4}) of tetrahedron.\Label{algline:gs-vtx}
    \li Retrieve vertex coordinates of tetrahedron ($4\ \times$ \texttt{vec4}).
    \li Determine side (\texttt{int}) of each vertex with respect to the hyperplane: \texttt{s0}, \texttt{s1}, \texttt{s2}, \texttt{s3}.
    \li Compute the result code \texttt{r = s0 + s1 << 1 + s2 << 2 + s3 << 3}.
    \li Use the look-up table in Fig.~\ref{fig:vis} to determine intersected edges.
    \li Compute the intersection point $\vec{p}$ for each intersected edge using \texttt{r}, along with any
    \zi necessary varyings for each emitted vertex (triangle normal, altitude).
    \li Project each intersection point $\vec{p}$ to the basis of the hyperplane.
    \li Compute output triangle coordinates $\vec{p}^\prime$ by transforming each $\vec{p}$
    \zi by the usual model-view-projection matrix.
  \end{algcode}
  \caption{Algorithm to compute the intersection region for a single tetrahedron within a geometry shader.
  Here, it is assumed that the input is a tetrahedron index which is the invocation ID of the shader.}
  \label{alg:vis-geometry-shader}%
\end{algsimple}

\subsection{Rendering $4d$ tetrahedra without a geometry shader.}
\label{sec:vis-vs}
As noted by Maunoury and Loseille~\cite{Maunoury_2022}, the injection of a geometry shader into the rendering pipeline may diminish the resulting frame rate of the visualization.
Furthermore, one of the shader optimization techniques suggested by Halladay consists of moving computation to a vertex shader~\cite{Halladay_2019}.
As a result, a solution that involves calculating intersections directly in the vertex shader will now be presented.
For a tetrahedral mesh with $n_t$ tetrahedra, notice that there will be at most $2n_t$ triangles and $6n_t$ vertices that need to be rendered.
The rendering pipeline can be invoked (using \texttt{glDrawArrays}) with a \emph{count} equal to $6n_t$ (using \texttt{GL\_TRIANGLES} as the \emph{mode}).
This spawns a vertex shader invocation for each vertex ($6n_t$ invocations in total), and the algorithm described in Section~\ref{sec:visualization} can be used to determine whether this vertex is part of an intersection primitive to be rasterized.
This procedure is outlined in Alg.~\ref{alg:vis-vertex-shader}.
The invocation ID encodes (1) which tetrahedron and (2) which triangle vertex (of the intersection primitive) are being processed - these can be extracted using modular arithmetic on the invocation ID.
The intersection algorithm then proceeds as in the previous section, outputting $3d$ coordinates in clip space.
Another look-up table is then used to determine the ``opposite" vertices in each possible rendering primitive.
The latter is necessary to calculate the altitude of the output vertex for the solid wireframe model used to visualize the boundary of the intersection primitives.

\begin{algsimple}[H]
  \scriptsize
  \begin{algcode}[processVertex]
    \zi \Inputs index of vertex in triangle (\texttt{gl\_VertexID}), hyperplane definition: $\vec{n}$ (\texttt{vec4}), $\vec{c}$ (\texttt{vec4})
    \zi \Outputs transformed vertex coordinates (in $3d$)

    \li Determine which tetrahedron is being processed: \texttt{tet = gl\_VertexID / 6}.
    \li Determine which vertex in the triangle is being processed: \texttt{vtx = gl\_VertexID -  6 * tet}.
    \li Retrieve vertex indices (\texttt{ivec4}) of tetrahedron \texttt{tet}.\Label{algline:gs-vtx}
    \li Retrieve vertex coordinates of tetrahedron ($4\ \times$ \texttt{vec4}).
    \li Determine side (\texttt{int}) of each vertex with respect to the hyperplane: \texttt{s0}, \texttt{s1}, \texttt{s2}, \texttt{s3}.
    \li Compute the result code \texttt{r = s0 + s1 << 1 + s2 << 2 + s3 << 3}.
    \li Determine which edge is intersected using a look-up table: \texttt{e = v2e[6 * shape + vtx]}
    \zi \Comment{// const int v2e[18] = int[](0, 0, 0, 0, 0, 0, 0, 1, 2, 0, 0, 0, 0, 1, 2, 1, 3, 2);}
    \zi \Comment{// \texttt{shape} is 0 (no intersection), 1 (triangle) or (2) quad, found using a look-up table on \texttt{r}.}
    \li Compute the intersection point $\vec{p}$ for the intersected edge using \texttt{r}, along with any
    \zi necessary varyings for each emitted vertex (triangle normal, position in triangle).
    \li Project the intersection point $\vec{p}$ to the basis of the hyperplane.
    \li Compute output coordinates $\vec{p}^\prime$ by transforming $\vec{p}$ by the usual model-view-projection matrix.
    \li When drawing the wireframe of the intersections (borders of the intersection regions), also
    \zi look up the opposite vertices in the triangle to determine the altitude of the processed vertex.
  \end{algcode}
  \caption{Algorithm to compute the clip space coordinates of a vertex in a rendering triangle (or quadrilateral) if the corresponding tetrahedron intersects the user-defined hyperplane.}
  \label{alg:vis-vertex-shader}%
\end{algsimple}

\subsection{Rendering $4d$ triangles with a geometry shader.}

The renderings presented in Section~\ref{sec:results} also contain thick red lines representing the CAD Edges.
Recall that each CAD Edge is associated with a triangulation from Section~\ref{sec:tessellate}.
These red lines are thus computed as the intersection of the used-defined hyperplane with each triangle associated with the CAD Edges (saved in the final output mesh that contains the tetrahedralizations).
The intersection between a $4d$ hyperplane with a triangle in $4d$ is a line, and a geometry shader can be used to calculate these line rendering primitives.
The efficiency of this approach is not explored since the Edge triangulations are typically much smaller than the tetrahedralizations.

\subsection{Rendering pentatopes.}

In the experiments of Section~\ref{sec:results}, the performance of the aforementioned algorithms with pentatopal meshes will also be evaluated.
Within a pentatope mesh, each pentatope is bounded by five tetrahedra, so these can be extracted within a single invocation of the geometry shader, and then passed to the same $4d$ tetrahedron visualization algorithm described above.
Since the tessellation algorithm described in Section~\ref{sec:tessellate} only produces tetrahedra, readily available pentatopal meshes produced by \texttt{avro}~\cite{Caplan_2020_CAD} are used to study the interactive performance when rendering pentatopes.

\section{Numerical Experiments}
\label{sec:results}

The meshing algorithm described in Section~\ref{sec:tessellate} will now be demonstrated.
The resulting meshes will then be used to evaluate the performance of the visualization system described in Section~\ref{sec:visualization}.
All meshes are saved in the \texttt{libMeshb} format~\cite{Marechal_2024}, which was extended to handle $4d$ coordinates as well as pentatopes.

The meshing algorithm was implemented in \texttt{C++}, using \texttt{EGADS} as the geometry backend.
The visualization algorithms were implemented using \texttt{OpenGL} with either \texttt{Core OpenGL} (Apple) or \texttt{EGL} (Linux) backends, building upon the \texttt{wings} framework~\cite{Caplan_2023}.

\subsection{Verification with analytic geometries.}

To begin, the volume of the tetrahedral meshes obtained from the tessellation algorithm are compared with the expected volume for analytically-defined moving geometries.
In particular, the geometry of a sphere and a torus will be studied.
It is important to note that, despite these geometries having analytical definitions, they are still represented as solid bodies in the \texttt{EGADS} framework.
As a result, it is possible that the results are limited by geometry tolerancing internally used by \texttt{OpenCASCADE}~\cite{RuizGirones_2016, RuizGirones_2021}.

All geometries are contained within a cube with sides of length $\ell$, similar to how a far-field would be added in a realistic simulation.
For these cases, the moving geometry traces a total volume ($v$) equal to the sum of the volume traced by the six faces of the cube plus the volume traced by the interior geometry ($v_i$):

\begin{equation}\label{eq:volume}
  v = 6\ell^2(t_f - t_0)  + v_i + v_{t_0} + v_{t_f},
\end{equation}

where $t_0$ and $t_f$ are the initial and final times of the geometry movement, respectively.
Without loss of generality, the entire motion is parametrized to be within the interval $[0, 1]$, so $t_0 = 0$ and $t_f = 1$.
The volumes $v_{t_0}$ and $v_{t_f}$ are the volumes of the initial and final tetrahedralizations (at $t_0$ and $t_f$, respectively).
The volume traced by the interior geometry will be described in each analytic case below.

\begin{figure}[H]
  \centering
  \hspace{-60pt}
  \includegraphics[width=\textwidth, trim={0 1.25cm 0 0.5cm}]{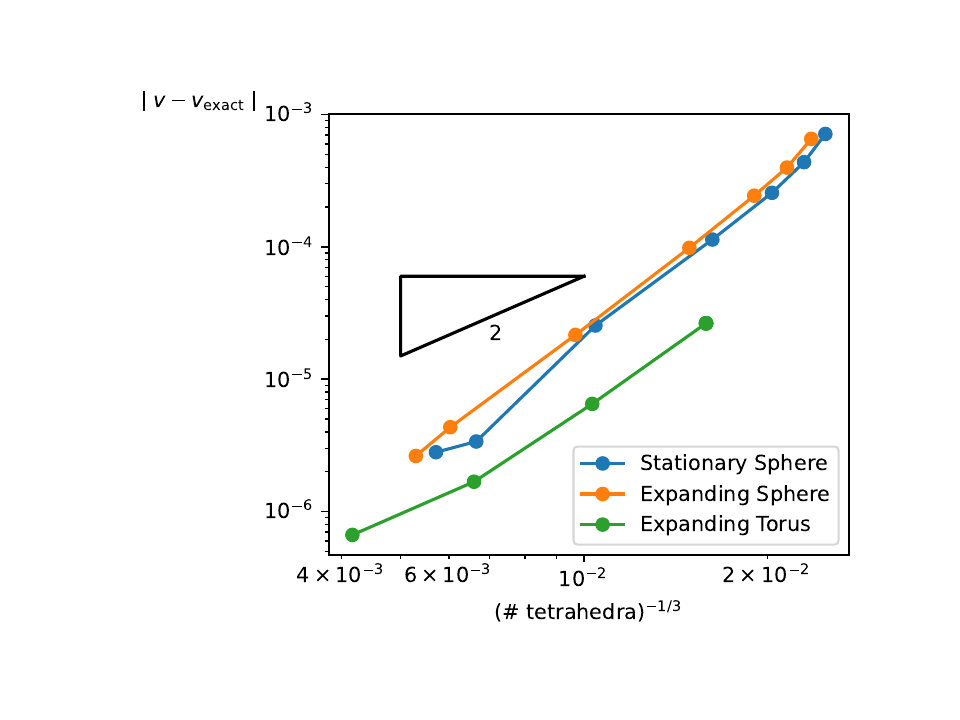}
  \caption{Convergence of the volume error for analytic geometries.}
  \label{fig:analytic-volume}
\end{figure}

Meshes were computed for varying resolutions of the geometry using $10$ time slabs for each case.
Once the final meshes were computed, the total volume of the tetrahedra was evaluated using the method described by Kahan~\cite{Kahan_2001}.
Sample visualizations are provided in the test case descriptions below.
Fig.~\ref{fig:analytic-volume} summarizes the results for the analytic geometries considered here, which shows that the error between the computed volume and the expected volume converges close to the expected second-order rate for straight-sided tetrahedra~\cite{Anderson_2023}.

\paragraph{Expanding sphere}

This test case considers a sphere of radius $r_0 = 0.1$ contained within a cube with a unit side length ($\ell = 1$).
The sphere linearly expands to a final radius of $r_f = 0.125$ at $t_f = 1$.
This linearly expanding sphere traces the geometry of a hypercone in $4d$, specifically a truncated hypercone.
Thus the interior volume ($v_{i}$) is the volume of this truncated hypercone, which can be derived as:

\begin{equation}\nonumber
  v_{i} = v_{\mathrm{hcs}}(r_f, a + 1) - v_{\mathrm{hcs}}(r_0, a), \quad \mbox{where}\ \ v_{\mathrm{hcs}}(r, h) = \frac{4\pi r^2}{3} \sqrt{r^2 + h^2},
\end{equation}
where $a = r_0 / (r_f - r_0)$ is the height of the portion of the hypercone that is truncated.
The subscript ``$\mathrm{hcs}$" denotes a hypercone with a spherical base.
The volumes at the initial and final hyperplanes are $v_{t_0} = \ell^3 - \frac{4}{3}\pi r_0^3$ and $v_{t_f} = \ell^3 - \frac{4}{3}\pi r_f^3$, respectively.

The case of a static sphere (no expansion) is also considered, which simply traces the geometry of a hypercylinder, so $v_i = 4\pi r_0^2 (t_f - t_0)$ and the volumes at the initial and final hyperplanes are $v_{t_0} = v_{t_f} = \ell^3 - \frac{4}{3}\pi r_0^3$.

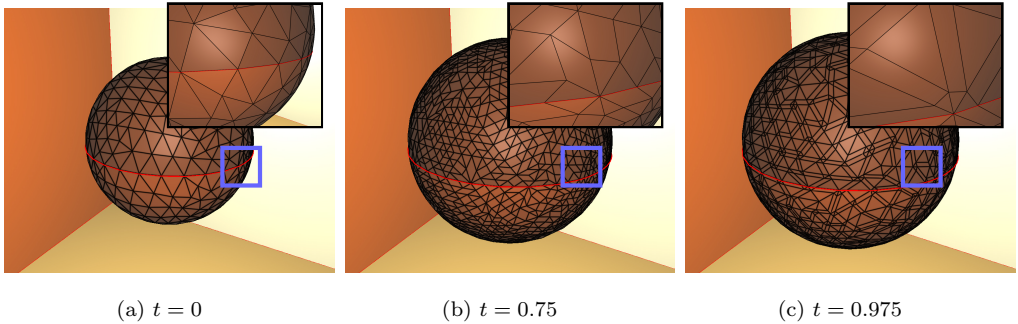
\begin{figure}[H]
  \centering
  \begin{subfigure}{0.32\textwidth}
    \begin{tikzpicture}
      \node at (0, 0) {\includegraphics[width=\textwidth]{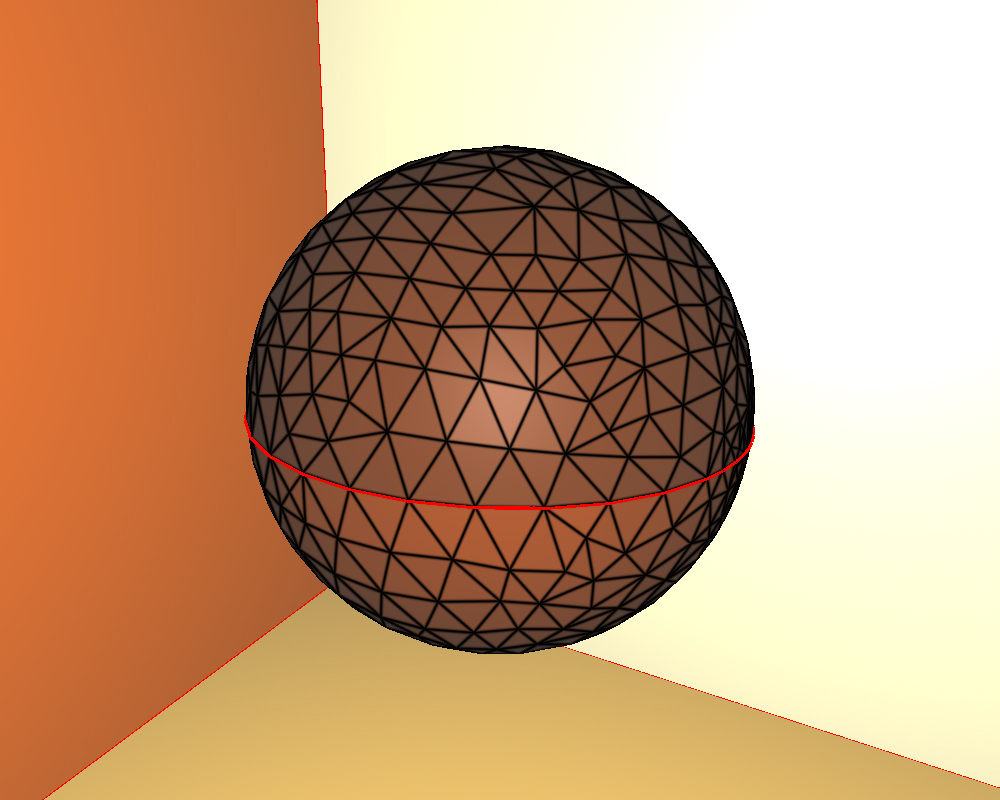}};
      \node [draw=black, text width=2cm, inner sep=0.01cm, ultra thick] at (1, 1) {\includegraphics[width=\textwidth]{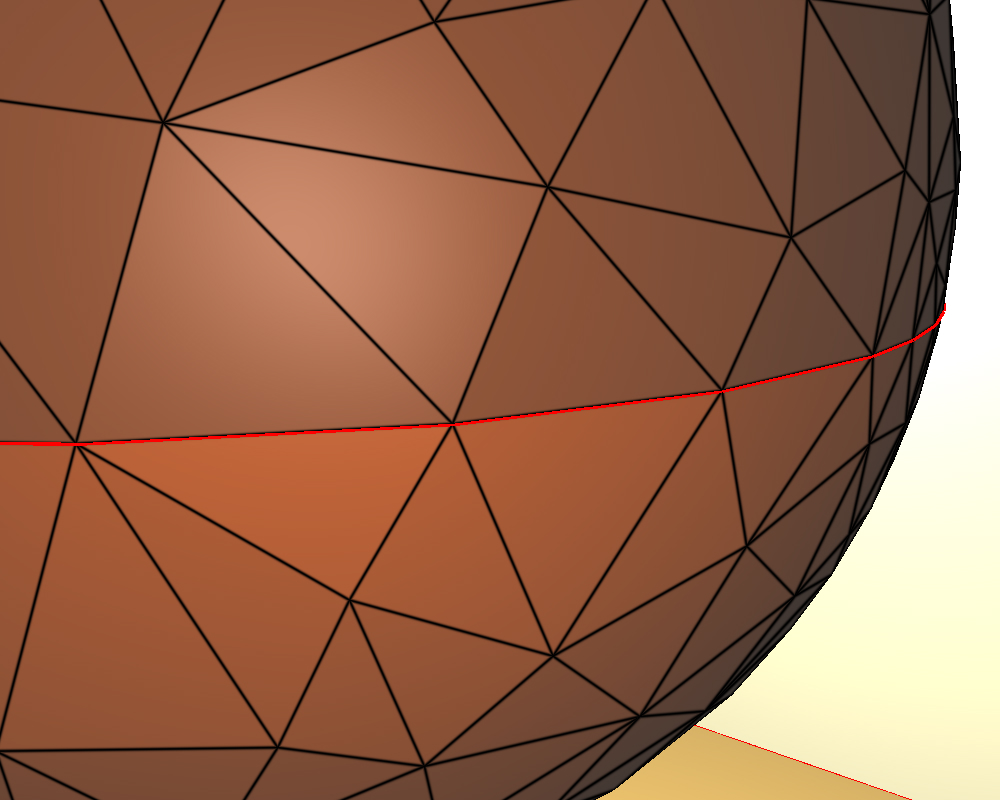}};
      \draw [draw=blue!60, ultra thick] (0.7, -0.6) rectangle ++(0.5, 0.5);
    \end{tikzpicture}
    \caption{$t = 0$}
  \end{subfigure}
  \begin{subfigure}{0.32\textwidth}
    \begin{tikzpicture}
      \node at (0, 0) {\includegraphics[width=\textwidth]{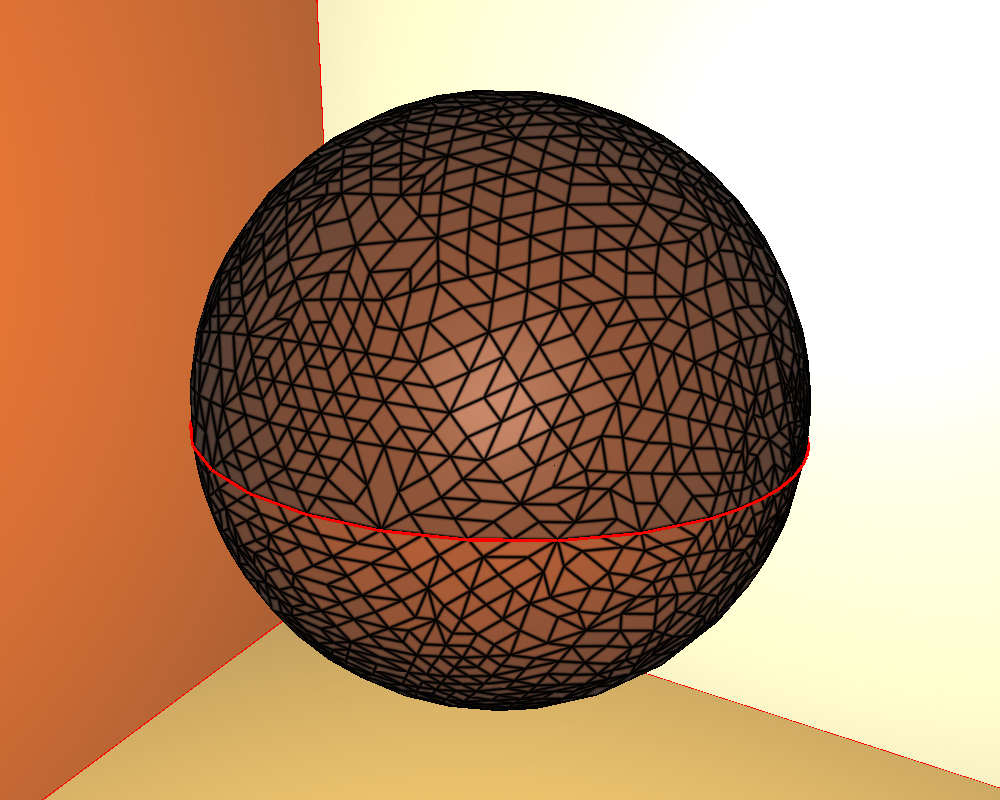}};
      \node [draw=black, text width=2cm, inner sep=0.01cm, ultra thick] at (1, 1) {\includegraphics[width=\textwidth]{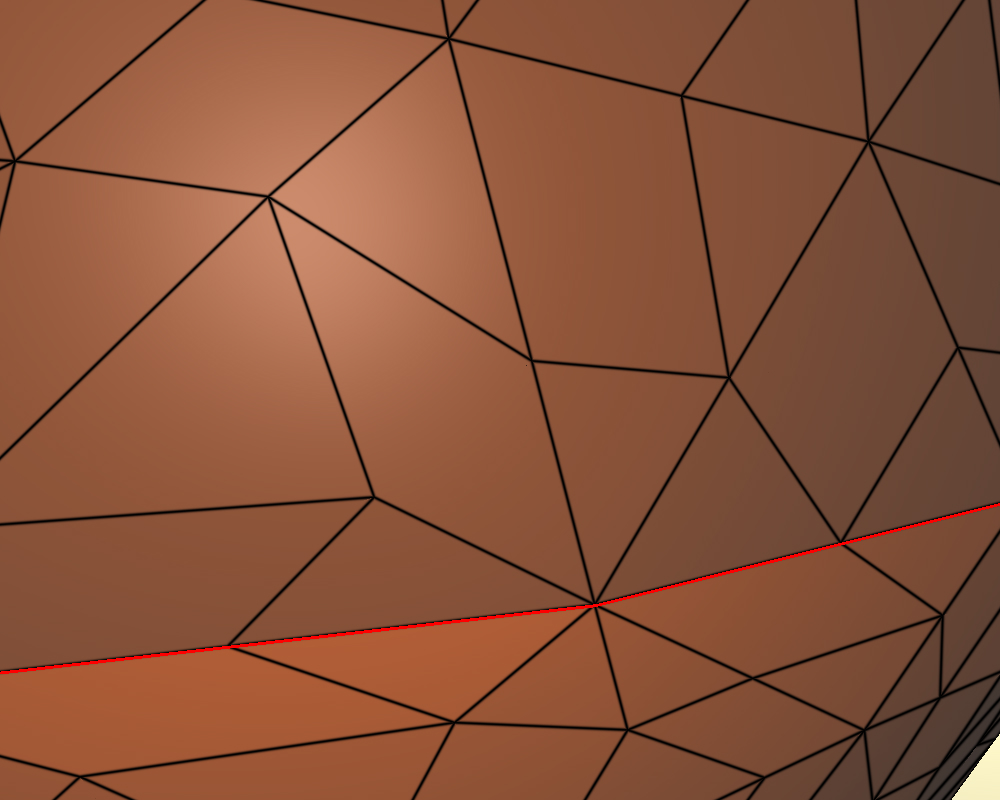}};
      \draw [draw=blue!60, ultra thick] (0.7, -0.6) rectangle ++(0.5, 0.5);
    \end{tikzpicture}
    \caption{$t = 0.75$}
  \end{subfigure}
  \begin{subfigure}{0.32\textwidth}
    \begin{tikzpicture}
      \node at (0, 0) {\includegraphics[width=\textwidth]{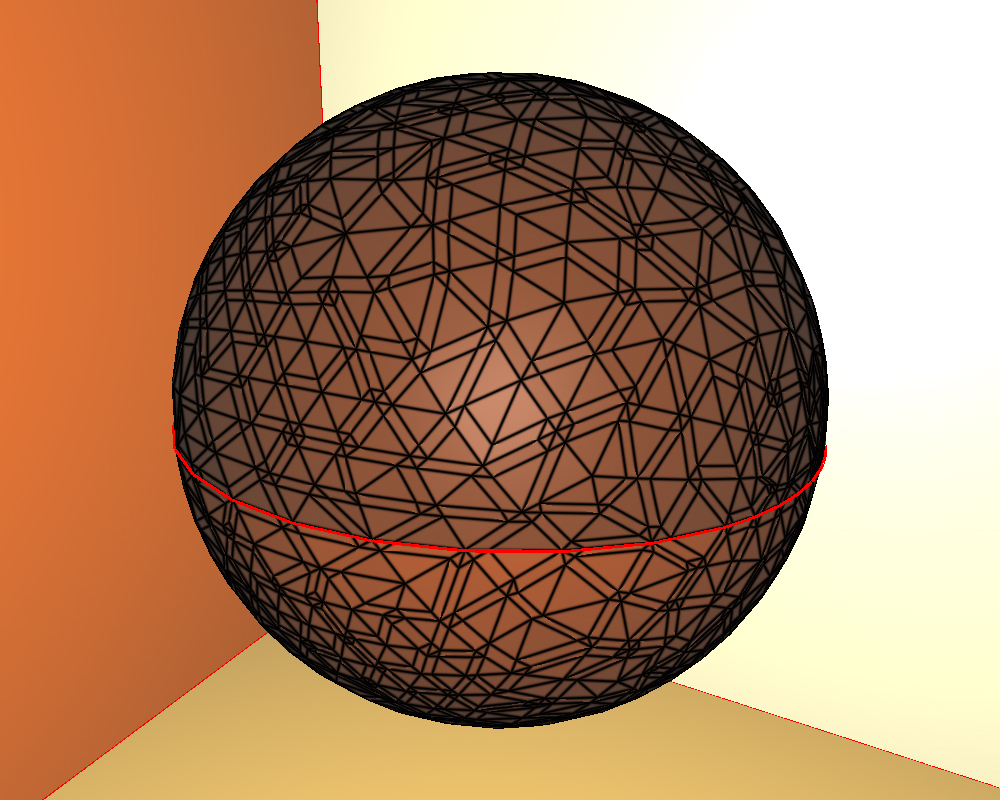}};
      \node [draw=black, text width=2cm, inner sep=0.01cm, ultra thick] at (1, 1) {\includegraphics[width=\textwidth]{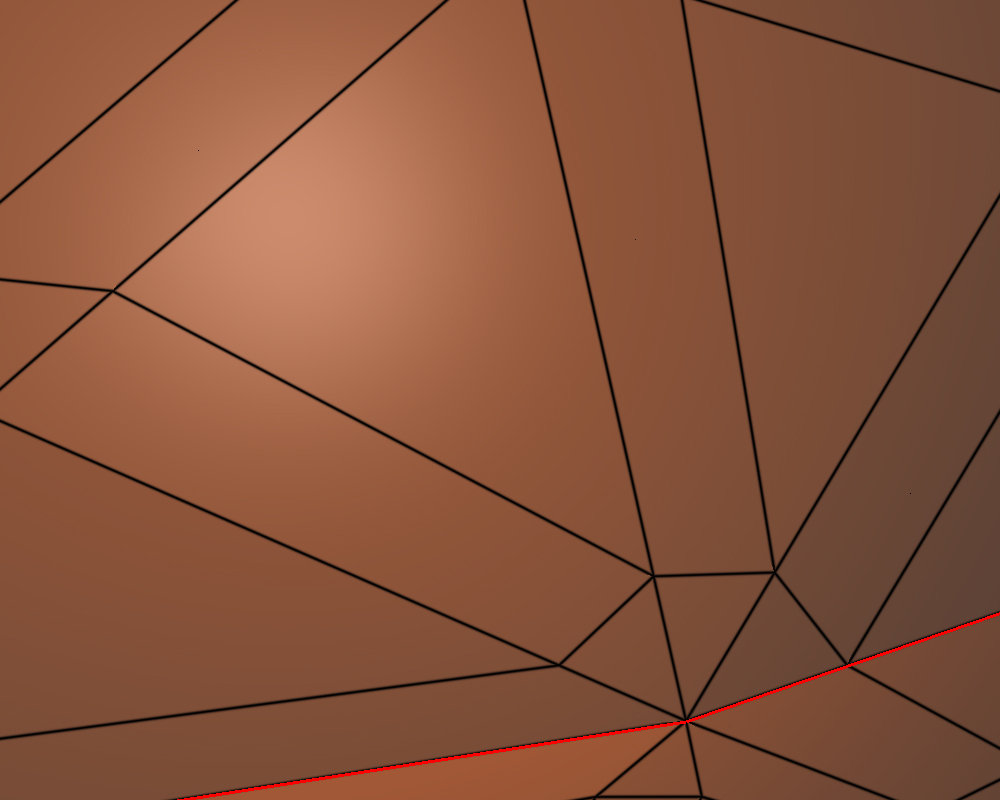}};
      \draw [draw=blue!60, ultra thick] (0.7, -0.6) rectangle ++(0.5, 0.5);
    \end{tikzpicture}
    \caption{$t = 0.975$}
  \end{subfigure}
  \caption{Visualization of the tetrahedral mesh for the expanding sphere case sliced at $t = 0$, $t = 0.75$ and $t = 0.975$.}
  \label{fig:expanding-sphere}
\end{figure}

Visualizations of the expanding sphere are provided in Fig.~\ref{fig:expanding-sphere}.
The algorithms described in Section~\ref{sec:visualization} were used to visualize sample tetrahedralizations produced by the algorithm of Section~\ref{sec:tessellate} at $t = 0$, $t = 0.75$ and $t = 0.975$.
At $t = 0$, the slice of the tetrahedral mesh only results into triangular intersection primitives, whereas both triangle and quadrilateral rendering primitives are apparent at $t = 0.75$ and $t = 0.975$.

It is worth mentioning that the tessellation parameters (passed to \texttt{EGADS}) were specified so as to maintain roughly the same edge lengths in the geometry tessellation as the sphere expands.
As a result, the number of vertices and triangles in the geometry tessellation varies between each time step in order to maintain the geometry resolution. 

\paragraph{Expanding torus}

This test case consists of an expanding torus with an initial tube (minor) radius of $r_{0} = 0.1$ and an initial major radius of $R_{0} = 0.4$.
The expanding torus is contained within a cube of side length $\ell = 1$.
A linear time-dependent scaling is applied to the torus such that it expands to a final tube radius of $r_{f} = 0.125$ and a final major radius of $R_{f} = 0.5$.
The analytic volume calculation follows a similar idea to the expanding sphere of the previous section, where the volume is that of a truncated hypercone.
The base of the cone is a $3d$ torus, so the interior volume traced by the expanding torus is

\begin{equation}
  v_{i} = v_{\mathrm{hct}}(r_{f}, R_{f}, a + 1) - v_{\mathrm{hct}}(r_{0}, R_{0}, a), \quad \mbox{where}\ \ v_{\mathrm{hct}}(r, R, h) = \frac{4\pi^2 r R}{3} \sqrt{R^2 + h^2},
\end{equation}
where $a = R_0 / (R_f - R_0)$.
The subscript ``$\mathrm{hct}$" denotes a hypercone with a toroidal base.
The volumes at the initial and final hyperplanes are $v_{t_0} = \ell^3 - 2\pi r_0^2 R_0$ and $v_{t_f} = \ell^3 - 2\pi r_f^2 R_f$, respectively.

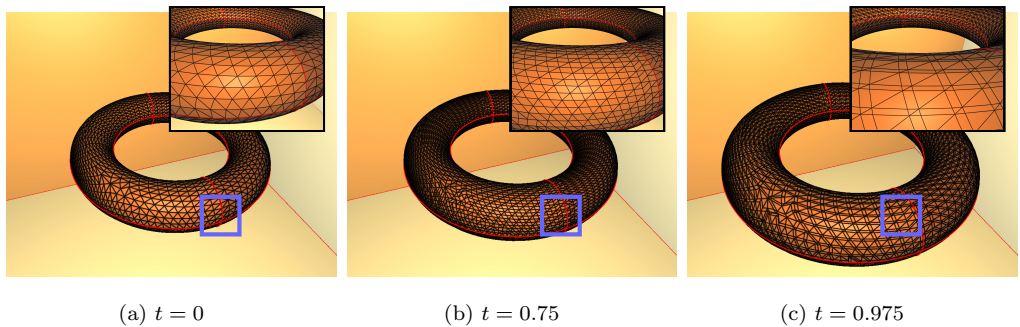
\begin{figure}[H]
  \centering
  \begin{subfigure}{0.32\textwidth}
    \begin{tikzpicture}
      \node at (0, 0) {\includegraphics[width=\textwidth]{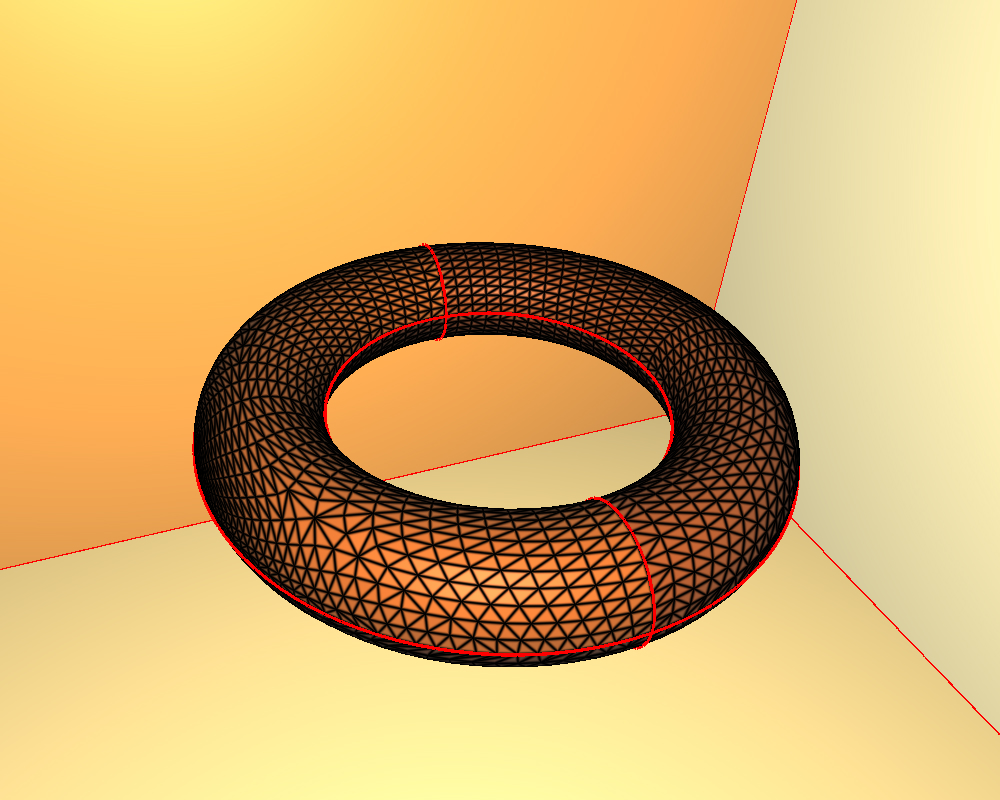}};
      \node [draw=black, text width=2cm, inner sep=0.01cm, ultra thick] at (1, 1) {\includegraphics[width=\textwidth]{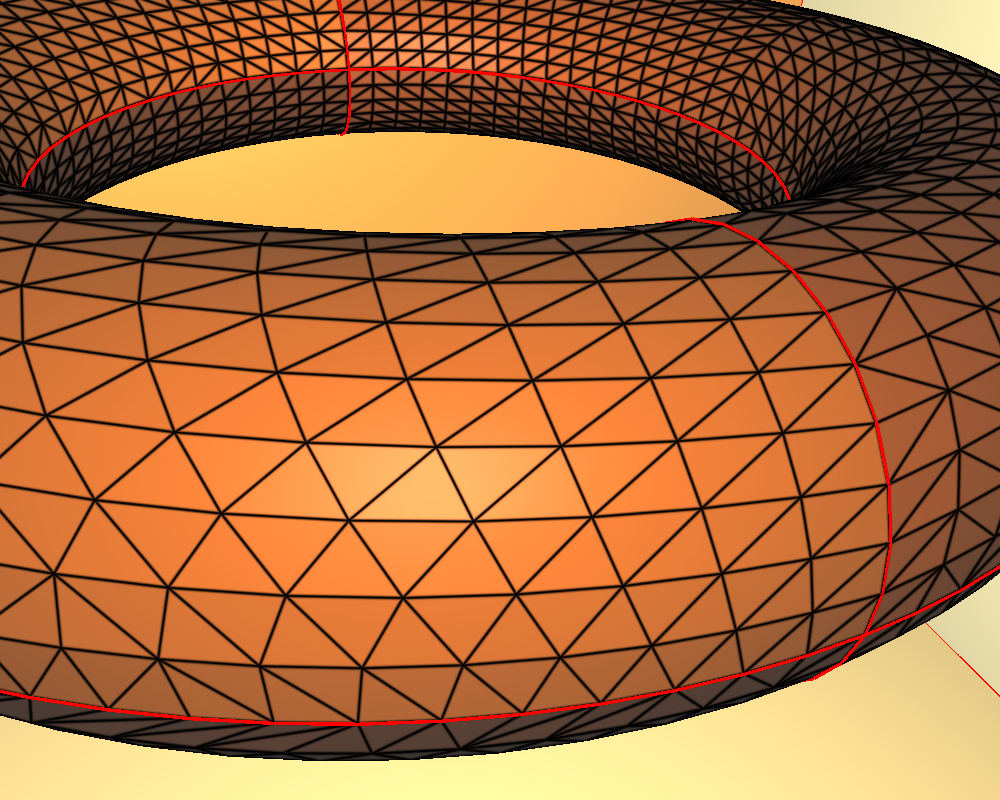}};
      \draw [draw=blue!60, ultra thick] (0.4, -1.2) rectangle ++(0.5, 0.5);
    \end{tikzpicture}
    \caption{$t = 0$}
  \end{subfigure}
  \begin{subfigure}{0.32\textwidth}
    \begin{tikzpicture}
      \node at (0, 0) {\includegraphics[width=\textwidth]{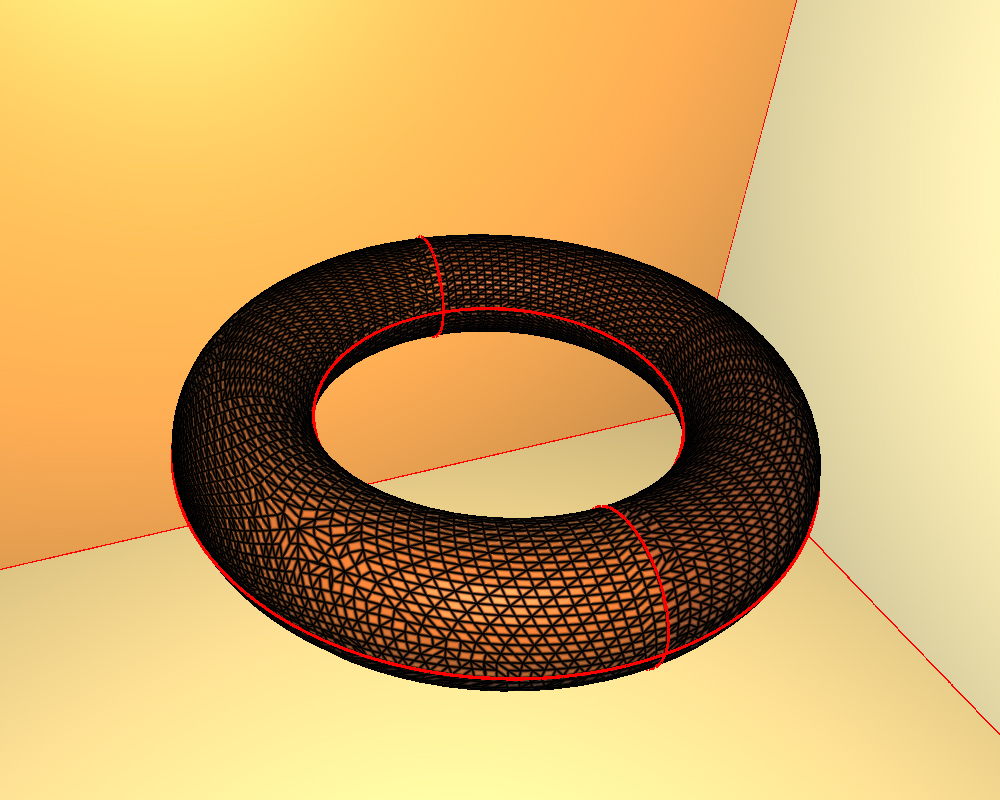}};
      \node [draw=black, text width=2cm, inner sep=0.01cm, ultra thick] at (1, 1) {\includegraphics[width=\textwidth]{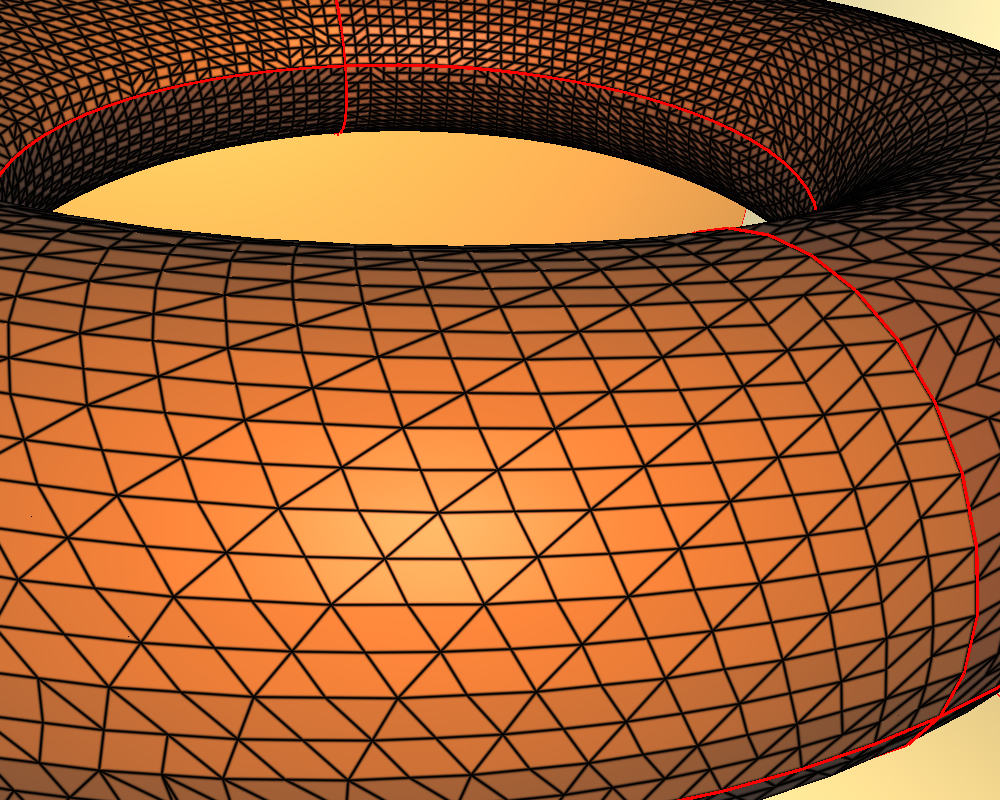}};
      \draw [draw=blue!60, ultra thick] (0.4, -1.2) rectangle ++(0.5, 0.5);
    \end{tikzpicture}
    \caption{$t = 0.75$}
  \end{subfigure}
  \begin{subfigure}{0.32\textwidth}
    \begin{tikzpicture}
      \node at (0, 0) {\includegraphics[width=\textwidth]{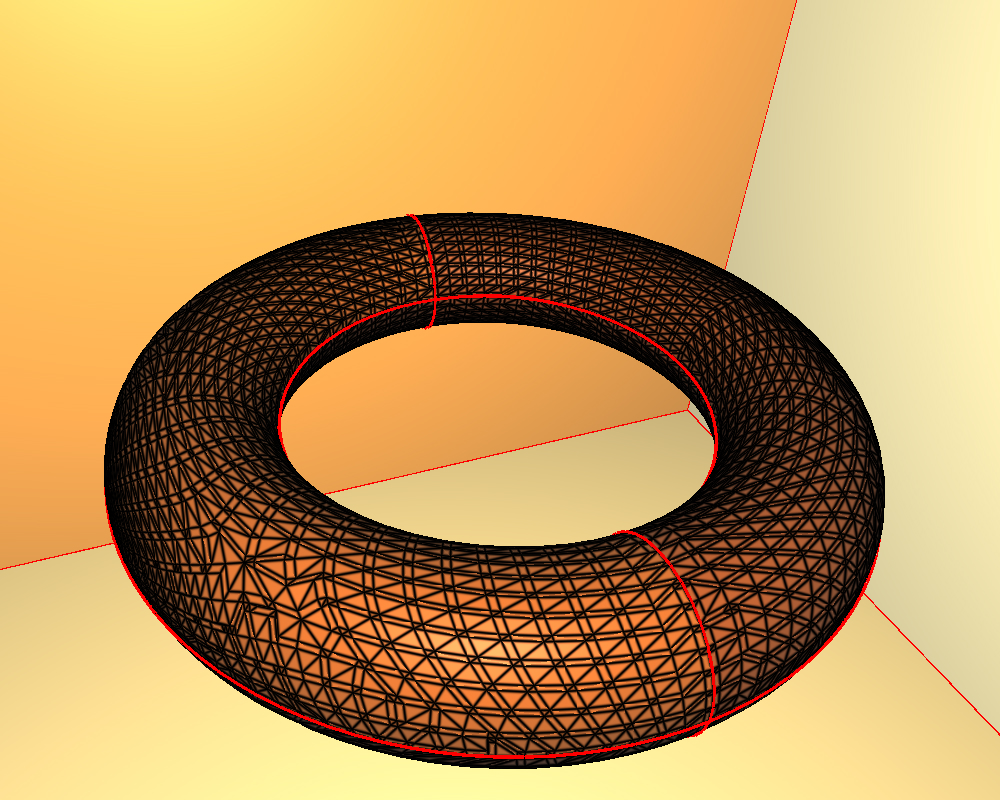}};
      \node [draw=black, text width=2cm, inner sep=0.01cm, ultra thick] at (1, 1) {\includegraphics[width=\textwidth]{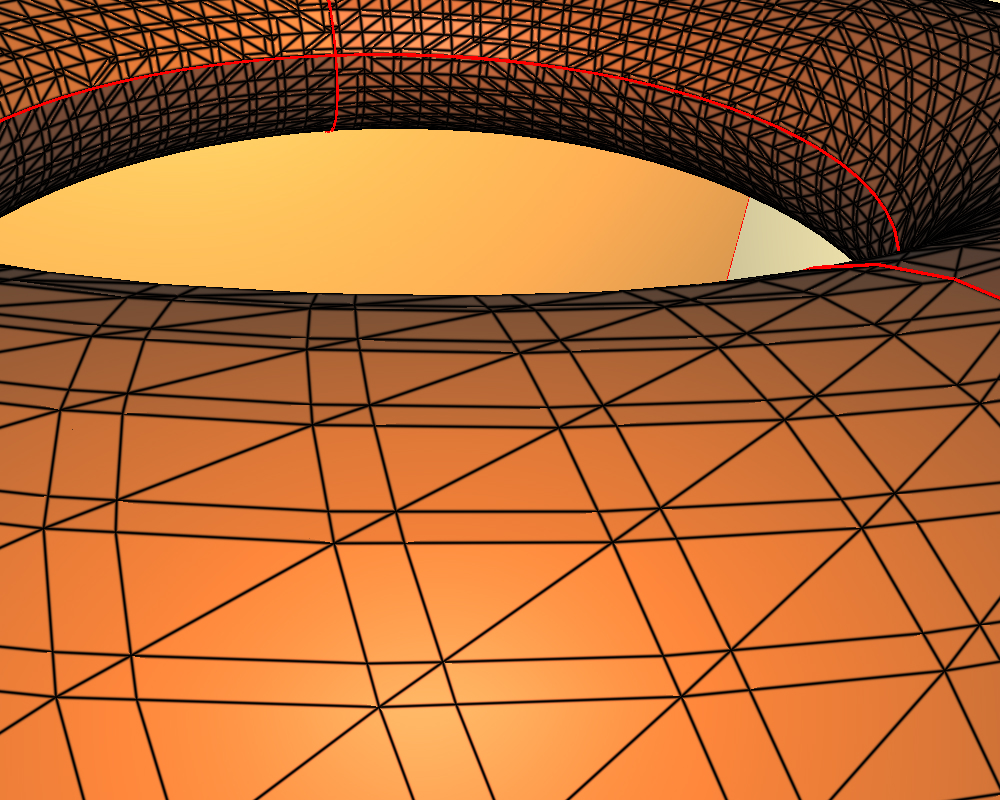}};
      \draw [draw=blue!60, ultra thick] (0.4, -1.2) rectangle ++(0.5, 0.5);
    \end{tikzpicture}
    \caption{$t = 0.975$}
  \end{subfigure}
  \caption{Visualization of the tetrahedral mesh for the expanding torus case sliced at at $t = 0$, $t = 0.75$ and $t = 0.975$.}
  \label{fig:expanding-torus}
\end{figure}

Sample visualizations of the expanding torus are provided in Fig.~\ref{fig:expanding-torus}.
Similar to the visualizations of the expanding sphere, the slice of the tetrahedral mesh at $t = 0$ only contains triangular intersection primitives, whereas both triangle and quadrilateral rendering primitives are apparent at $t = 0.75$ and $t = 0.975$.

\subsection{Application of the meshing algorithm to complex geometries.}
\label{sec:complex-geometries}

\paragraph{Wing-flap deployment}

Now the algorithms described in this paper will be used to create and visualize spacetime meshes of more complex geometries, beginning with the geometry illustrated in Fig.~\ref{fig:intro} in order to model a wing-flap configuration of an aircraft.
The wing and flap geometries were generated with \texttt{OpenCSM}~\cite{Dannenhoffer_2013} using \texttt{partspanflap1.csm} in the Engineering Sketch Pad \texttt{data} directory.
To define the geometry at each time step $t$, the flap body was rotated by an angle $\theta(t)$ about the $z$-axis (spanwise direction), centered at the point $(0.8, 0, 0)$, where $\theta(t) = -30 + 60(t/N)$ (degrees) and $N$ is the number of spacetime slabs.
A farfield parallelepiped was added to the geometry where the length of each side is a factor of $(10, 50, 2)$ in the $x$-, $y$- and $z$- directions, respectively, of the bounding box lengths of the wing.

\begin{figure}[H]
  \centering
  \def\pwidth{2.5cm}
  \begin{subfigure}{0.475\textwidth}
    \begin{tikzpicture}
      \node at (0, 0) {\includegraphics[width=\textwidth]{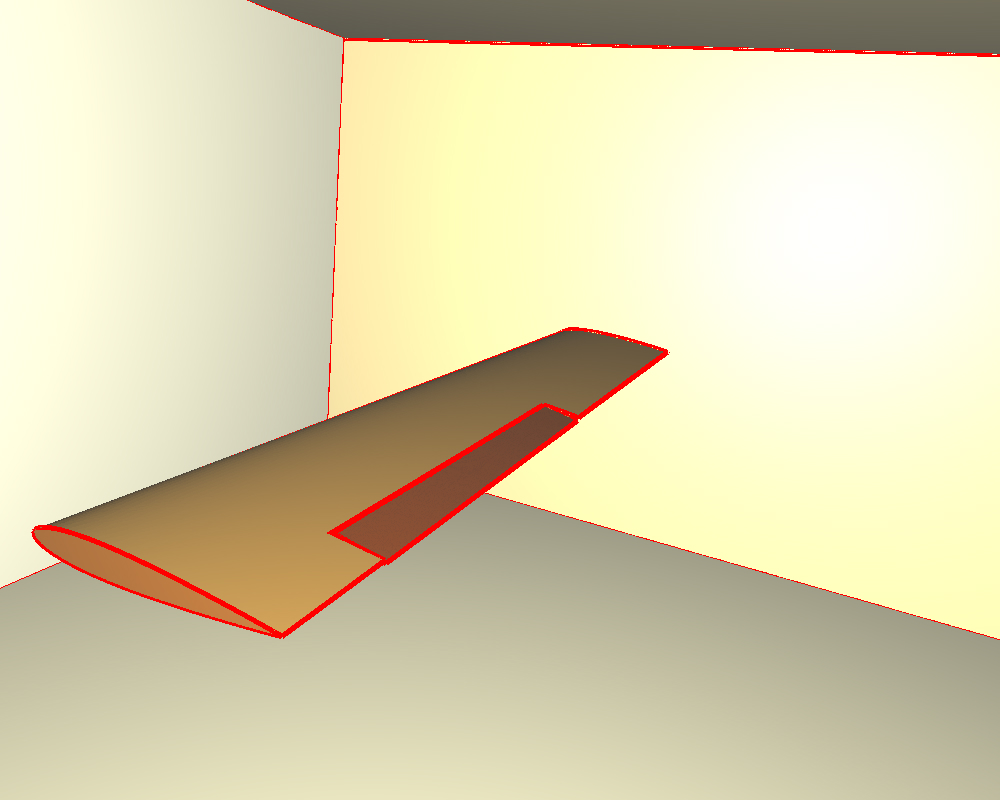}};
      \node (a) [draw=black, text width=\pwidth, inner sep=0.01cm, ultra thick] at (1.8cm, 1.55cm) {\includegraphics[width=\pwidth]{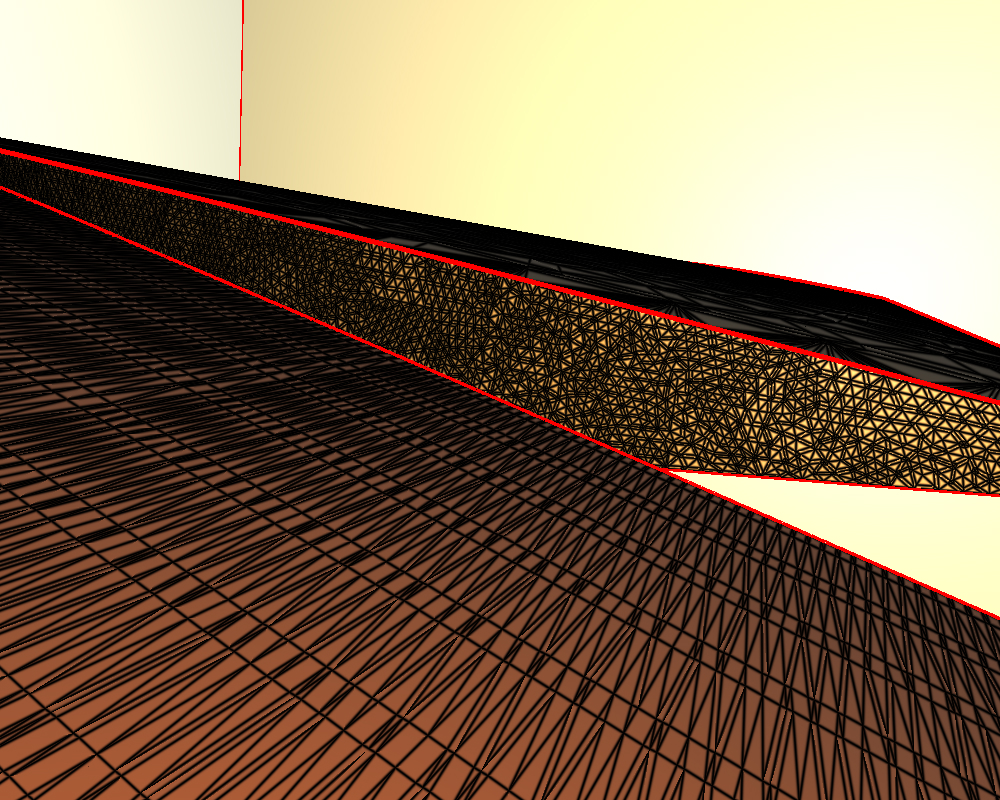}};
      \node (b) [draw=black, text width=\pwidth, inner sep=0.01cm, ultra thick] at (1.8, -1.55) {\includegraphics[width=\pwidth]{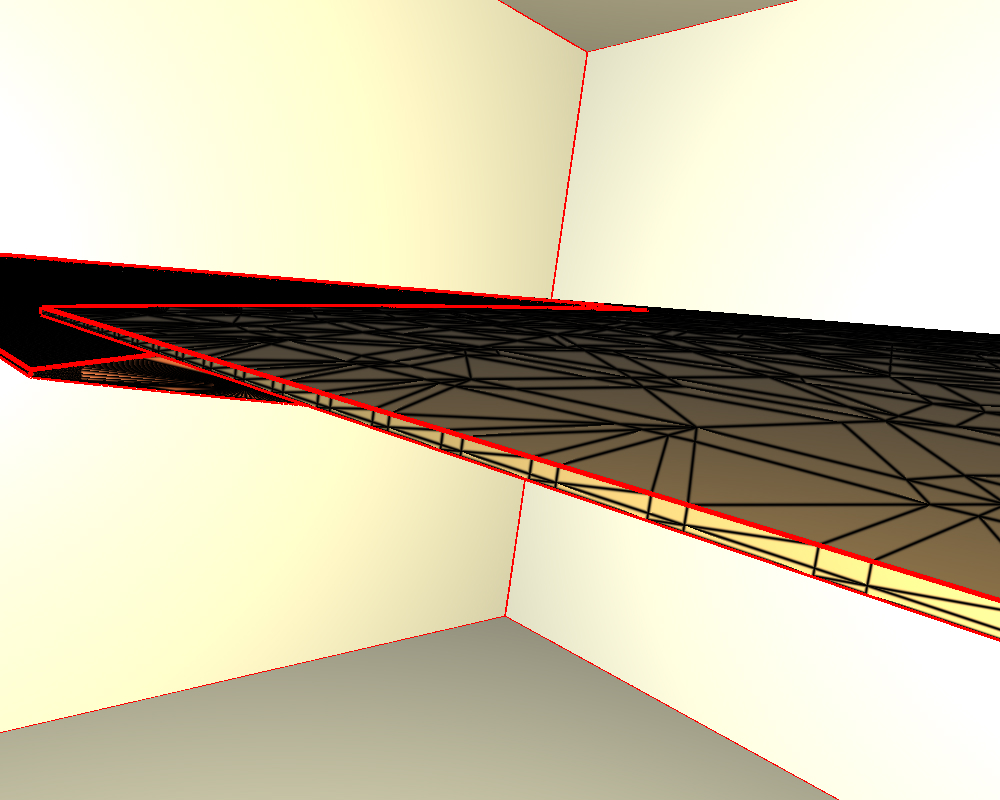}};
      \draw[ultra thick, ->, >=latex, blue!70] (b) to[out=85, in=-300] (0.5, 0);
      \draw[ultra thick, ->, >=latex, blue!70] (a.west) to[out=180, in=-170] (0.4, -0.1);
    \end{tikzpicture}
  \end{subfigure}
  \begin{subfigure}{0.475\textwidth}
    \begin{tikzpicture}
      \node at (0, 0) {\includegraphics[width=\textwidth]{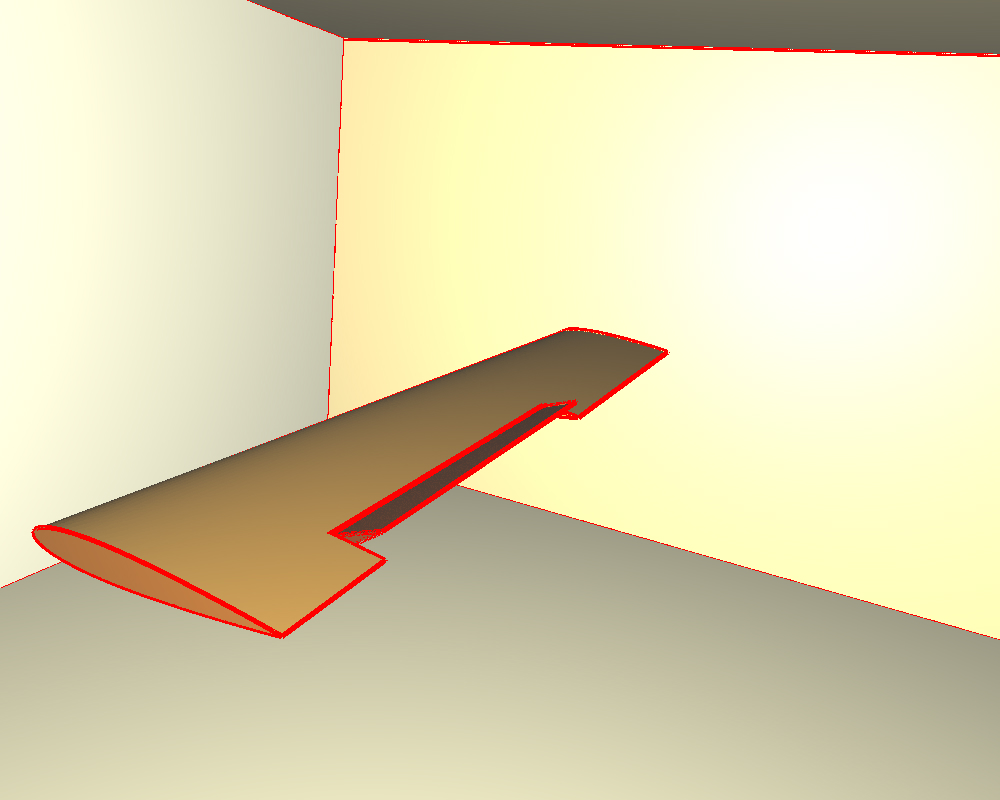}};
      \node (a) [draw=black, text width=\pwidth, inner sep=0.01cm, ultra thick] at (1.8cm, 1.55cm) {\includegraphics[width=\pwidth]{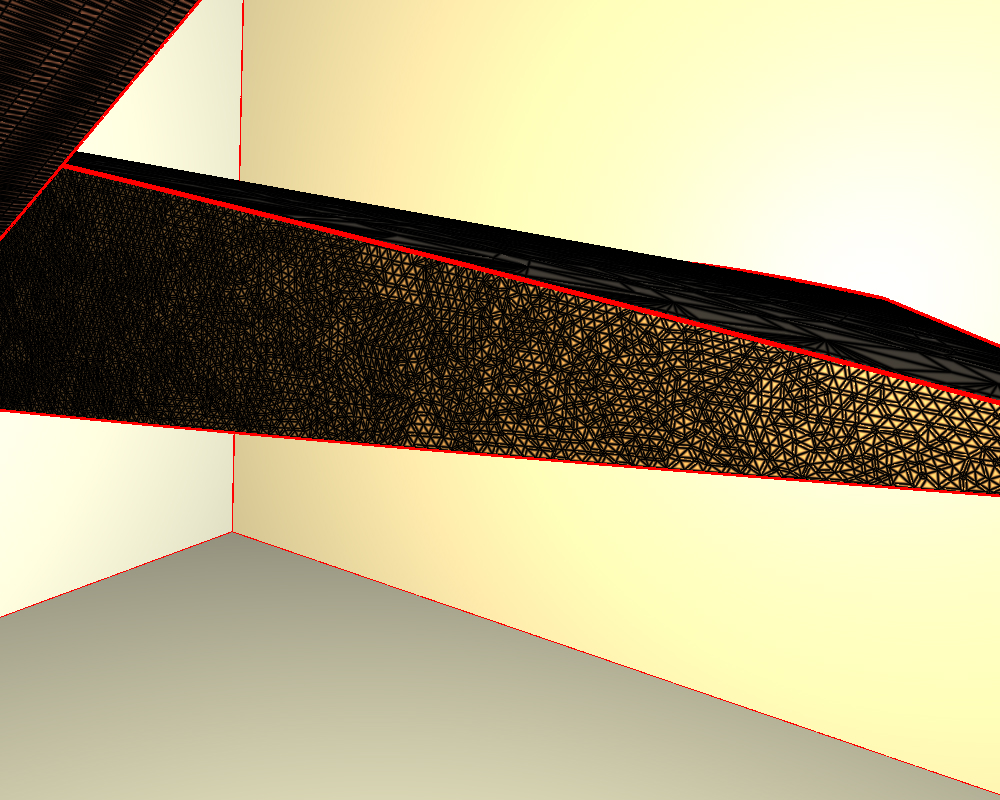}};
      \node (b) [draw=black, text width=\pwidth, inner sep=0.01cm, ultra thick] at (1.8, -1.55) {\includegraphics[width=\pwidth]{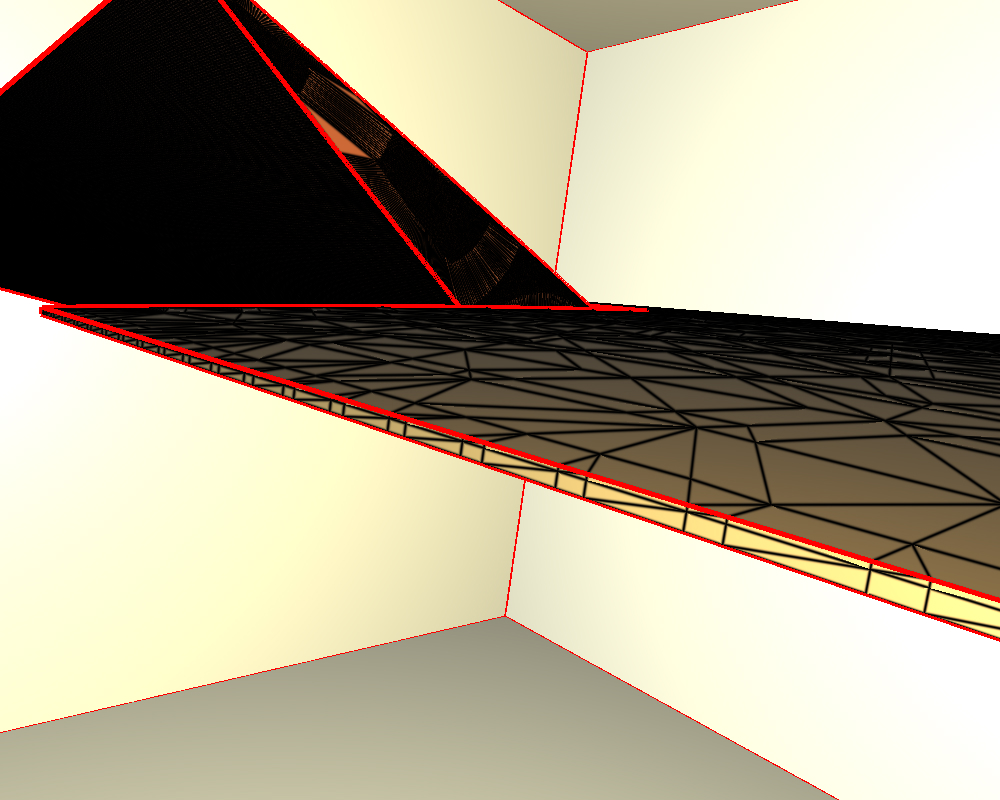}};
      \draw[ultra thick, ->, >=latex, blue!70] (b) to[out=85, in=-300] (0.5, 0);
      \draw[ultra thick, ->, >=latex, blue!70] (a.west) to[out=180, in=-170] (0.4, -0.1);
    \end{tikzpicture}
  \end{subfigure}
  \caption{Visualization of the wing-flap deployment at $t = 0.425$ and $t = 0.975$ seconds.}
  \label{fig:wingflap-vis}
\end{figure}
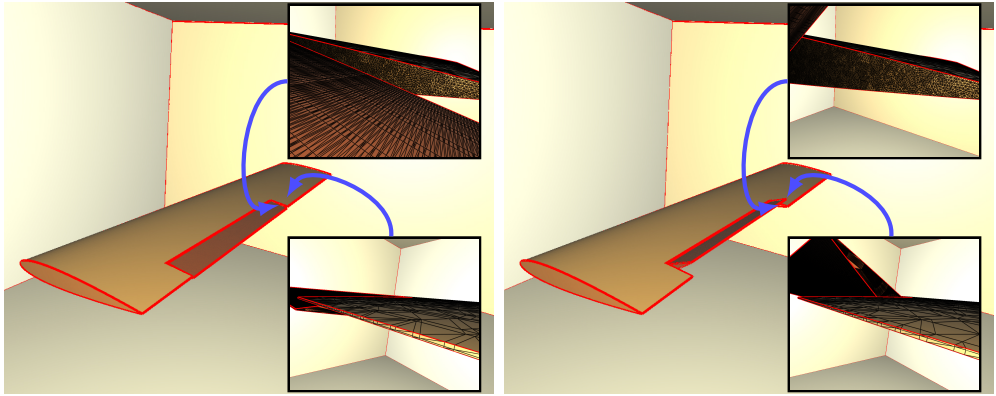

Sample visualizations of the sliced tetrahedral mesh (produced with the algorithms in Section~\ref{sec:visualization}) are shown in Fig.~\ref{fig:wingflap-vis}.

A total of 10 time slabs was used to create coarse- and fine-resolution meshes for this moving geometry, consisting of approximately 12M and 45M tetrahedra, respectively.
The resolution of the geometry for these two cases is controlled by the parameters passed to the \texttt{EGADS} tessellator at each time step.
Table~\ref{tab:wingflap-stats} summarizes the final mesh statistics for these meshes, as well as some timing data of the current algorithm (run on a 2021 10-core Apple M1 Pro).
Ultimately, very few Steiner vertices were inserted relative to the total number of vertices for each mesh: 46 in the case of the coarser mesh (of about 2M vertices) and 536 in the case of the finer mesh (of about 7.8M vertices).

\begin{table}[H]
  \centering
  \begin{tabular}{rcc}
    & Coarse & Fine \\ \hline
\# tetrahedra& 11,943,904& 45,070,546\\ \hline
\# vertices& 2,029,491& 7,780,008\\ \hline
total time (sec.)& 172& 874\\ \hline
initial/final tetrahedralization (sec.)& 11& 68\\ \hline
geometry construction (sec.)& 0.804& 0.815\\ \hline
geometry tessellation (sec.)& 57& 352\\ \hline
edge triangulation (sec.)& 0.002& 0.008\\ \hline
face tetrahedralization (sec.)& 92& 410\\ \hline
\# Steiner vertices& 46& 536\\ \hline

  \end{tabular}
  \caption{Mesh and timing statistics of the tessellation algorithm for the wing-flap geometry.}
  \label{tab:wingflap-stats}
\end{table}

\paragraph{Rotating wind turbine blades}

The developed tessellation and visualization algorithms were also applied to the wind turbine geometry illustrated in Fig.~\ref{fig:intro}.
The rotor was generated using the \texttt{prop1.csm} file (with 3 blades) in the \texttt{data/basic} directory of the Engineering Sketch Pad.
The cylindrical mast has a length of 20 (aligned with the $z$-direction), with a base of radius $r_m = 0.125$ centered on $(0.35, 0, 0)$.
A hub was added where the rotor meets the cylindrical mast, and the nose was modified to create a flat surface in order to circumvent issues when creating meshes near the nose tip.

\begin{figure}
  \centering
  \begin{tikzpicture}
    \node at (0, 0) {\includegraphics[width=0.6\textwidth]{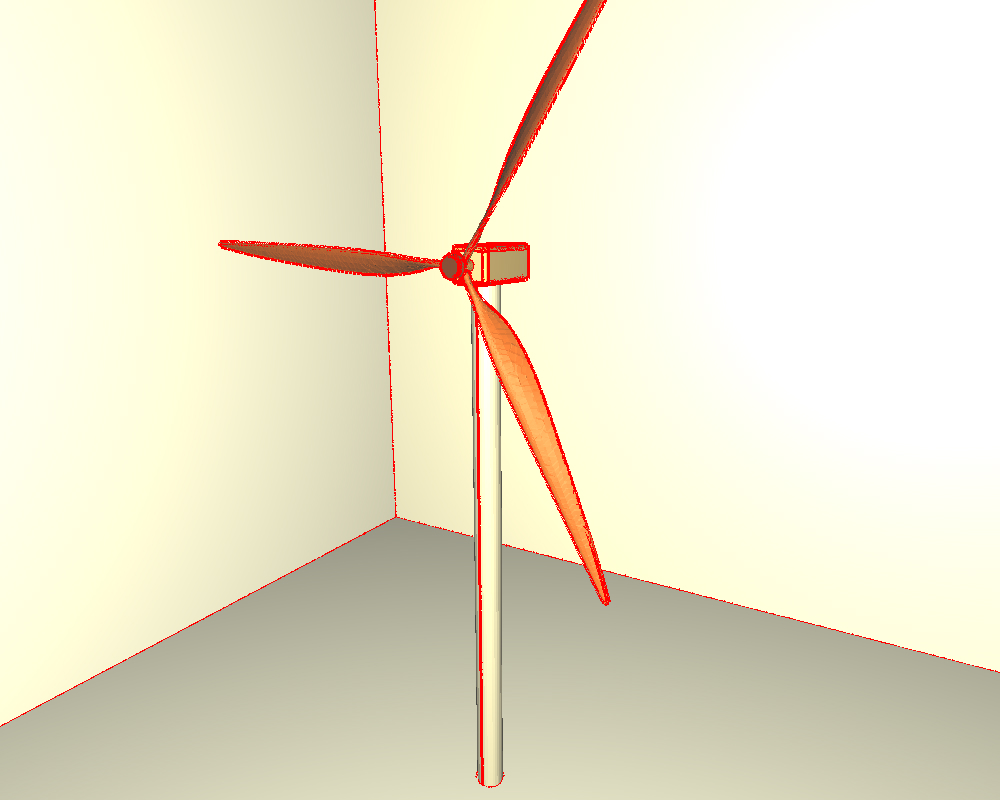}};
    \node [draw=black, text width=4cm, inner sep=0.01cm, ultra thick] at (3, 2) {\includegraphics[width=\textwidth]{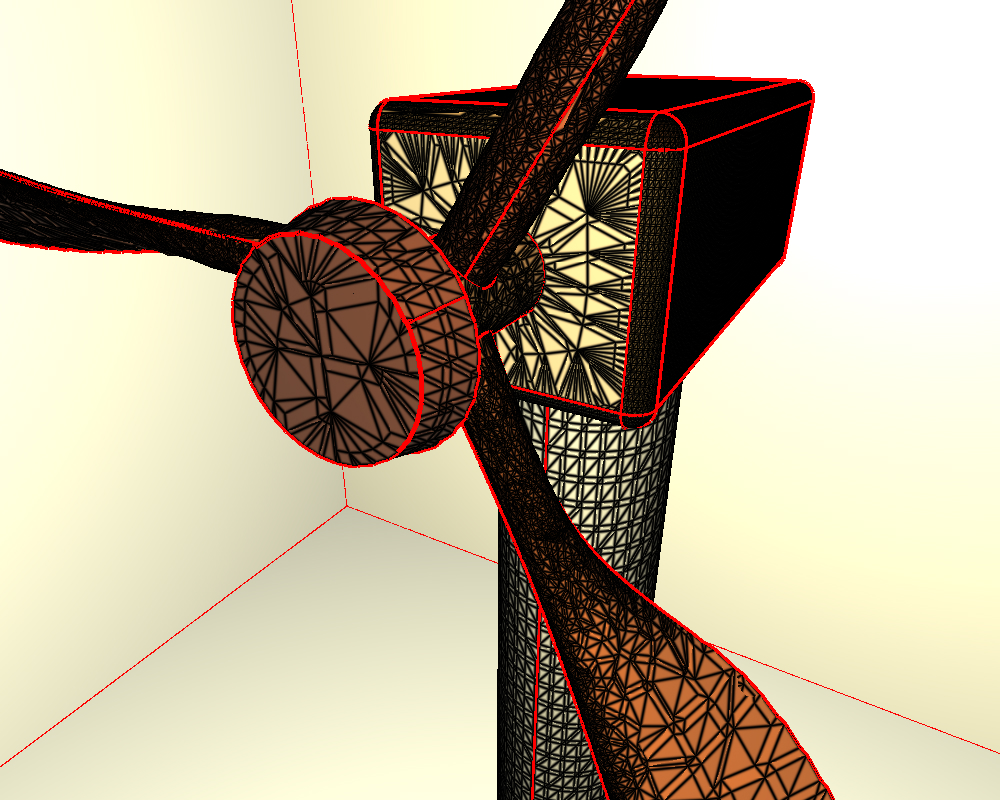}};
    \node [draw=black, text width=4cm, inner sep=0.01cm, ultra thick] at (3, -1.5) {\includegraphics[width=\textwidth]{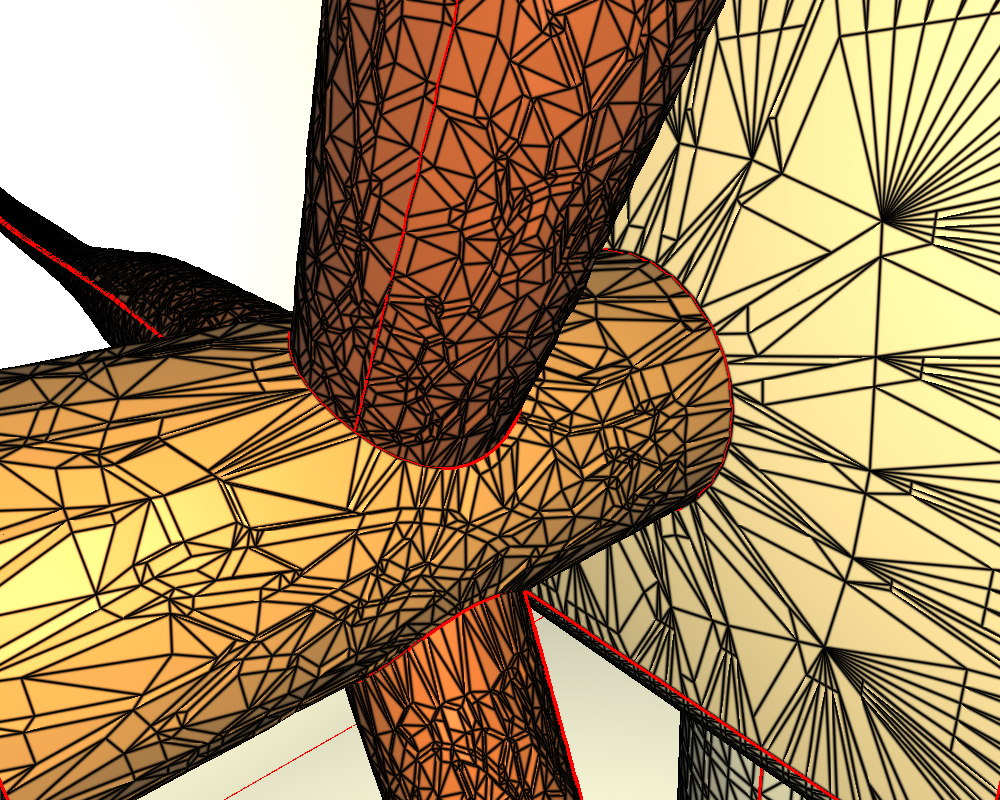}};
    \node [draw=black, text width=4cm, inner sep=0.01cm, ultra thick] at (-2.5, -1.5) {\includegraphics[width=\textwidth]{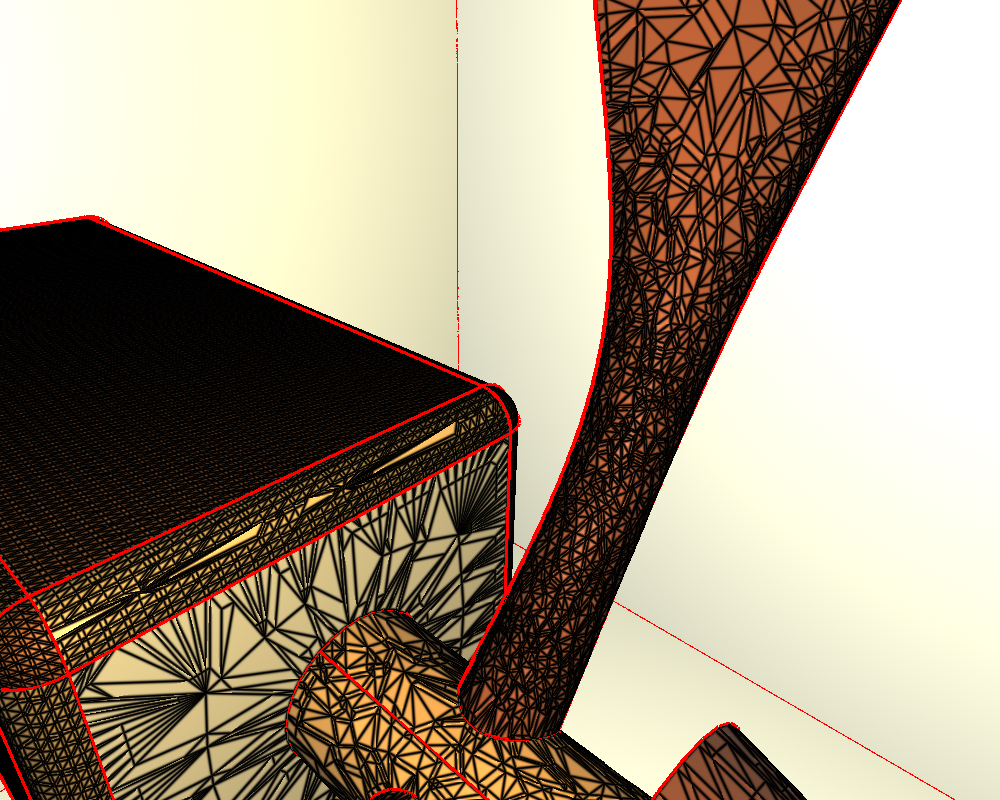}};
  \end{tikzpicture}
  \caption{Visualization of the wind turbine mesh at $t = 0.425$ seconds.}
  \label{fig:windturbine-vis-1}
\end{figure}

\begin{figure}
  \centering
  \begin{tikzpicture}
    \node at (0, 0) {\includegraphics[width=0.6\textwidth]{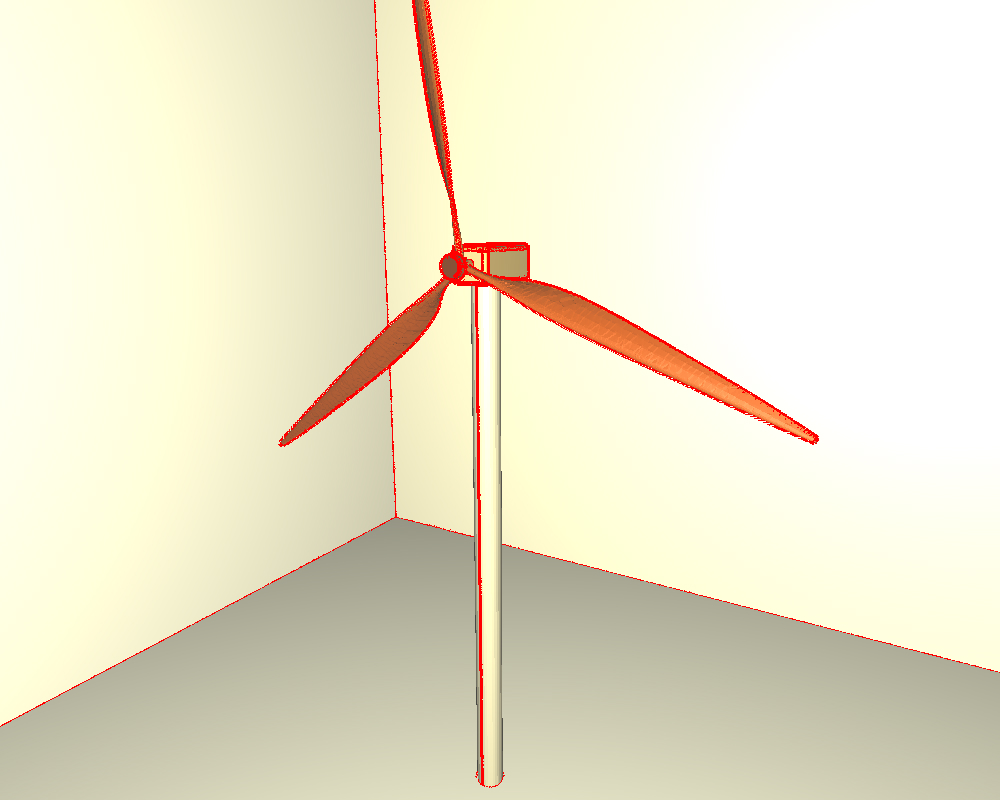}};
    \node [draw=black, text width=4cm, inner sep=0.01cm, ultra thick] at (3, 2) {\includegraphics[width=\textwidth]{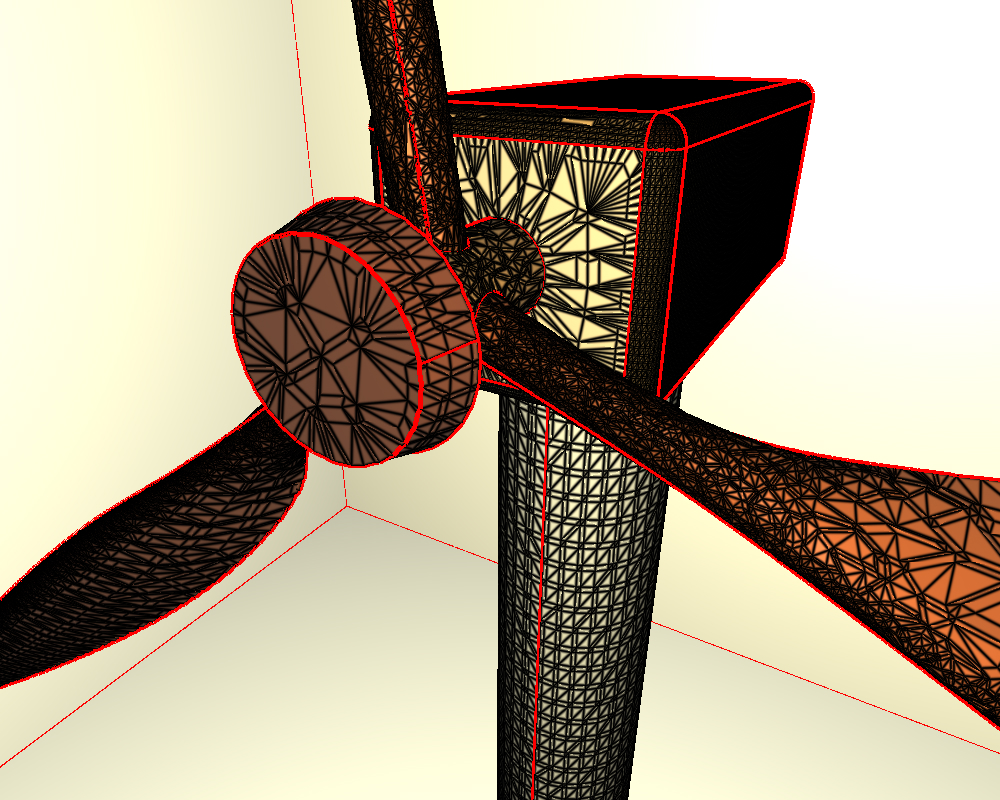}};
    \node [draw=black, text width=4cm, inner sep=0.01cm, ultra thick] at (3, -1.5) {\includegraphics[width=\textwidth]{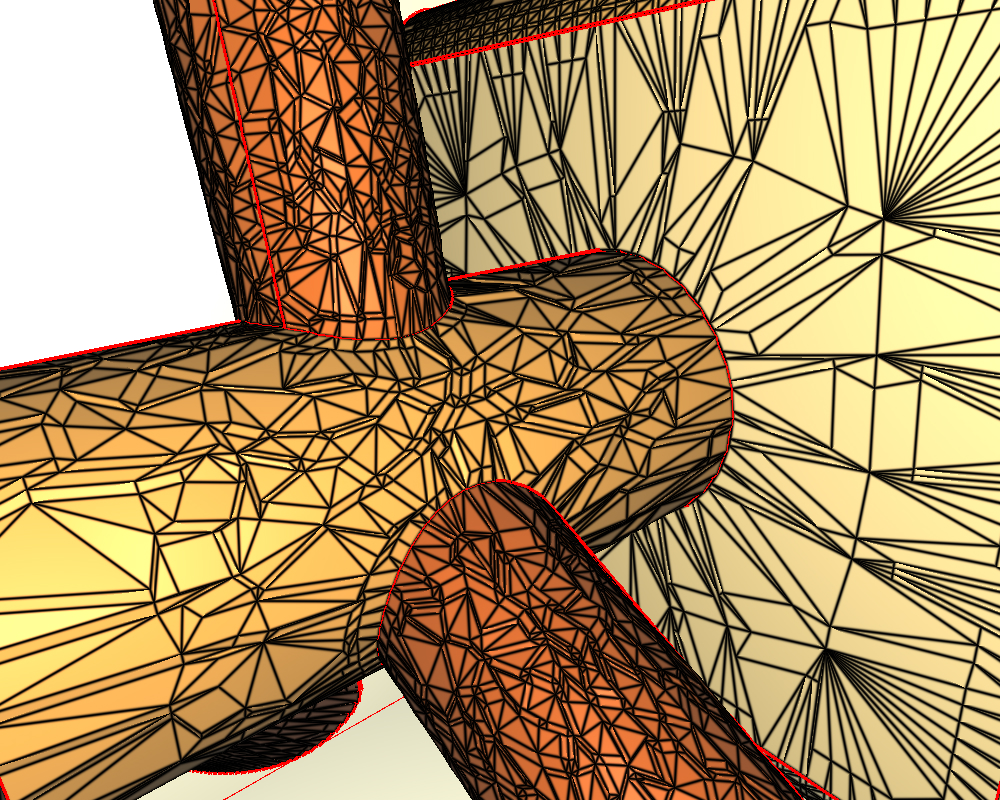}};
    \node [draw=black, text width=4cm, inner sep=0.01cm, ultra thick] at (-2.5, -1.5) {\includegraphics[width=\textwidth]{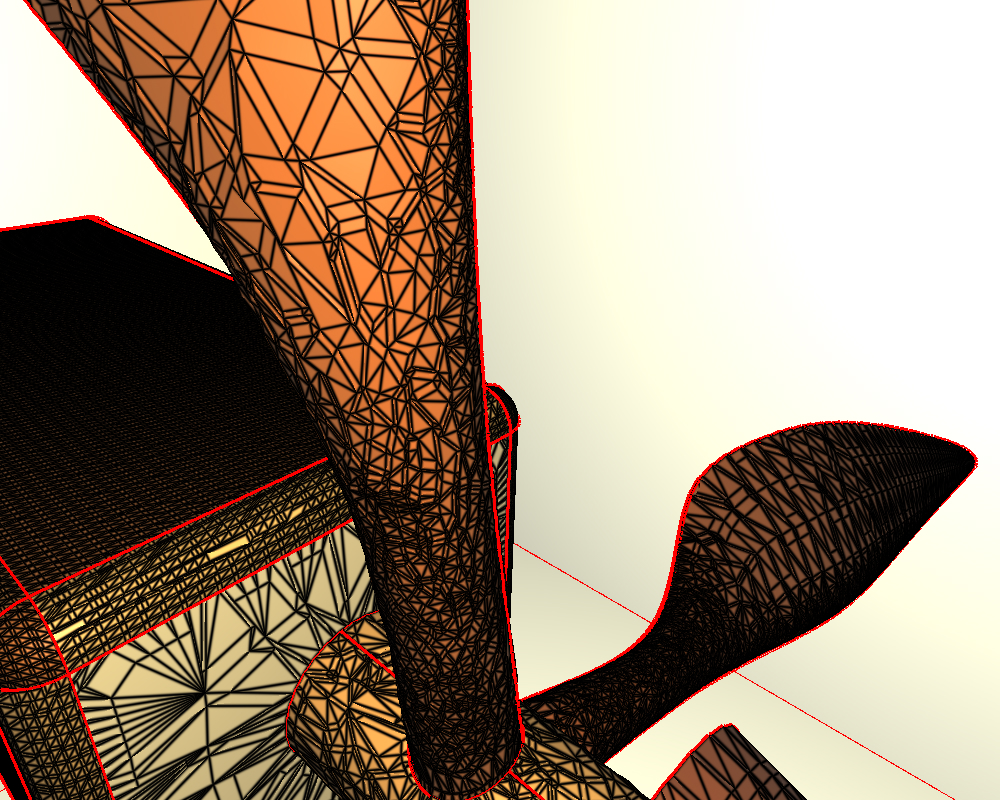}};
  \end{tikzpicture}
  \caption{Visualization of the wind turbine mesh at $t = 0.975$ seconds.}
  \label{fig:windturbine-vis-2}
\end{figure}

The entire rotor was linearly rotated by an angle of $360^\circ$ over the time interval $[0, 1]$ seconds.
A total of 30 time slabs were used to create coarse- and fine-resolution mesh for this geometry, which resulted in meshes with approximately 34M and 86M tetrahedra, respectively.
Again, the number of Steiner vertices (added in between time slabs) is relatively small compared to the total number of vertices, about 0.07\% of the vertices in the coarse-resolution mesh and about 0.02\% of the vertices for the finer mesh.
Table~\ref{tab:windturbine-stats} contains details of the final meshes, as well as the timing breakdown of the algorithm, which was also run on a 10-core Apple M1 Pro (2021).
Sample visualizations of the sliced tetrahedral meshes for this test case are shown in Figs.~\ref{fig:windturbine-vis-1} ($t = 0.425$) and \ref{fig:windturbine-vis-2} ($t = 0.975$).

\begin{table}[H]
  \centering
  \begin{tabular}{rcc}
    & Coarse & Fine \\ \hline
\# tetrahedra& 34,395,682& 85,849,469\\ \hline
\# vertices& 5,525,545& 14,086,607\\ \hline
total time (sec.)& 556& 2024\\ \hline
initial/final tetrahedralization (sec.)& 9.5& 18\\ \hline
geometry construction (sec.)& 6.4& 8.0\\ \hline
geometry tessellation (sec.)& 164& 1063\\ \hline
edge triangulation (sec.)& 0.001& 0.004\\ \hline
face tetrahedralization (sec.)& 275& 581\\ \hline
\# Steiner vertices& 3,918& 2,688\\ \hline

  \end{tabular}
  \caption{Mesh and timing statistics of the tessellation algorithm for the wind turbine geometry.}
  \label{tab:windturbine-stats}
\end{table}

\subsection{Performance of the visualization algorithms.}

Equipped with tetrahedral meshes of moving geometries, the performance of the visualization algorithm (of Section~\ref{sec:visualization}) will now be evaluated.
The average rendering time over 50 samples is used to measure the performance, where the scene parameters (model rotation, hyperplane slicing location) were defined randomly for each sample.
Triangle rendering (for the CAD Edges) was turned off for the analysis that follows in order to purely evaluate the rendering time for tetrahedra or pentatopes with the two approaches discussed in Section~\ref{sec:visualization}.
These two approaches are labelled ``GS" for the geometry-shader-based approach described in Section~\ref{sec:vis-gs} and Alg.~\ref{alg:vis-geometry-shader} and ``VS" for the vertex-shader-based approach described in Section~\ref{sec:vis-vs} and Alg.~\ref{alg:vis-vertex-shader}.

These approaches are evaluated using three GPUs: (1) that integrated with a 10-core Apple M1 Pro (2021), (2) an NVIDIA L4 GPU and (3) an NVIDIA P100 GPU (16 GB). 
Each sample was rendered off-screen to an $800 \times 600$ canvas, and the render time was measured using a \texttt{GL\_TIME\_ELAPSED} query over the invocation of the rendering pipeline.

\begin{figure}
  \centering
  \includegraphics[width=0.8\textwidth]{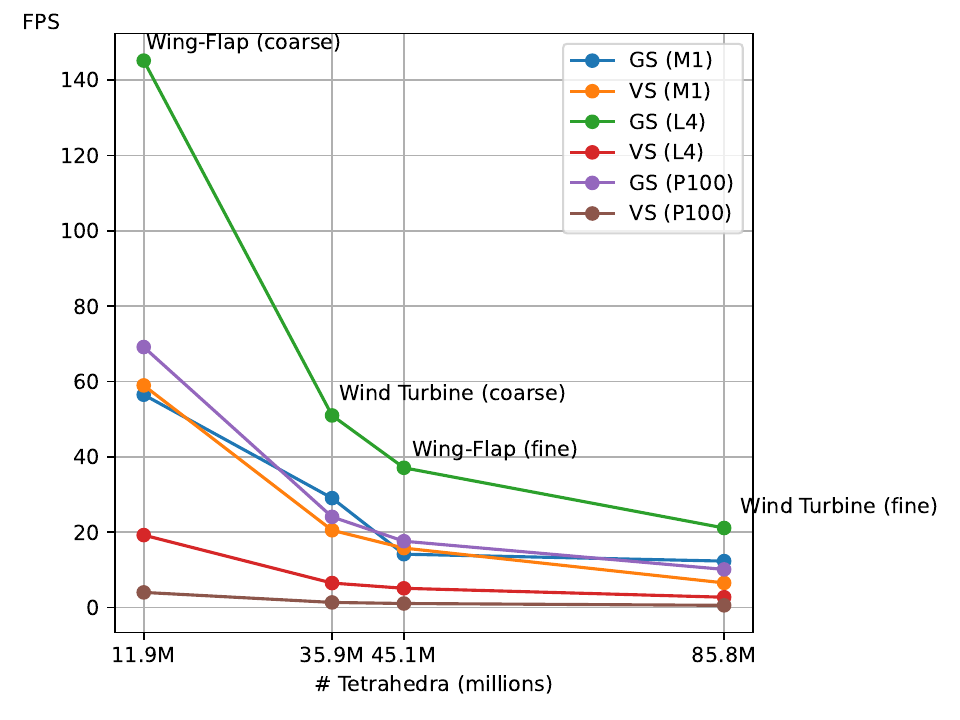}
  \caption{Frame rate obtained when rendering the meshes described in Section~\ref{sec:complex-geometries} with the geometry-shader- and vertex-shader-based approaches on three GPUs.}
  \label{fig:fps-complex-geometries}
\end{figure}

The frame rate obtained when rendering the tetrahedral meshes of Section~\ref{sec:complex-geometries} is shown in Fig.~\ref{fig:fps-complex-geometries}, again with the labels denoting the geometry-shader-based (GS) or vertex-shader-based (VS) approaches.
With the M1 GPU, the performance of both approaches is about the same, whereas the NVIDIA L4 GPU exhibits a higher frame rate (measured as the number of frames per second) when using the approach that includes a geometry shader.
With this approach, the frame rate is greater than 30 FPS for the 45.1M tetrahedra mesh, and drops to about 20 FPS for the 85.8M tetrahedra mesh.
A similar pattern is observed with the NVIDIA P100 GPU whereby higher frame rates are observed when using the geometry-shader-based algorithm.

\paragraph{Pentatopal meshes}

Since the visualization algorithm also supports rendering pentatopal meshes, the visualizer was also tested using meshes created by \texttt{avro}~\cite{Caplan_2019_PhD, Caplan_2020_CAD}.
Here, \texttt{avro} was used to generate an anisotropic $4d$ mesh within a $[0, 1]^4$ tesseract with a metric field designed to align with an expanding spherical wave.

\begin{figure}
  \centering
  \begin{subfigure}{0.32\textwidth}
    \includegraphics[width=\textwidth]{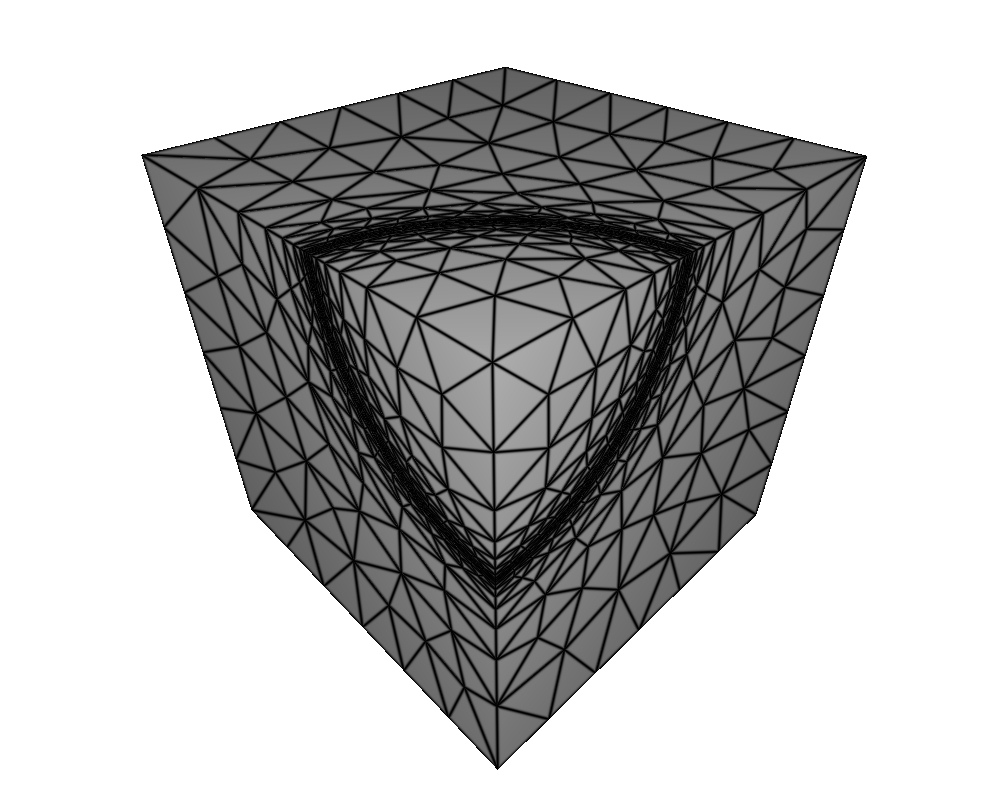}
    \caption{$t = 0$}
  \end{subfigure}
  \begin{subfigure}{0.32\textwidth}
    \includegraphics[width=\textwidth]{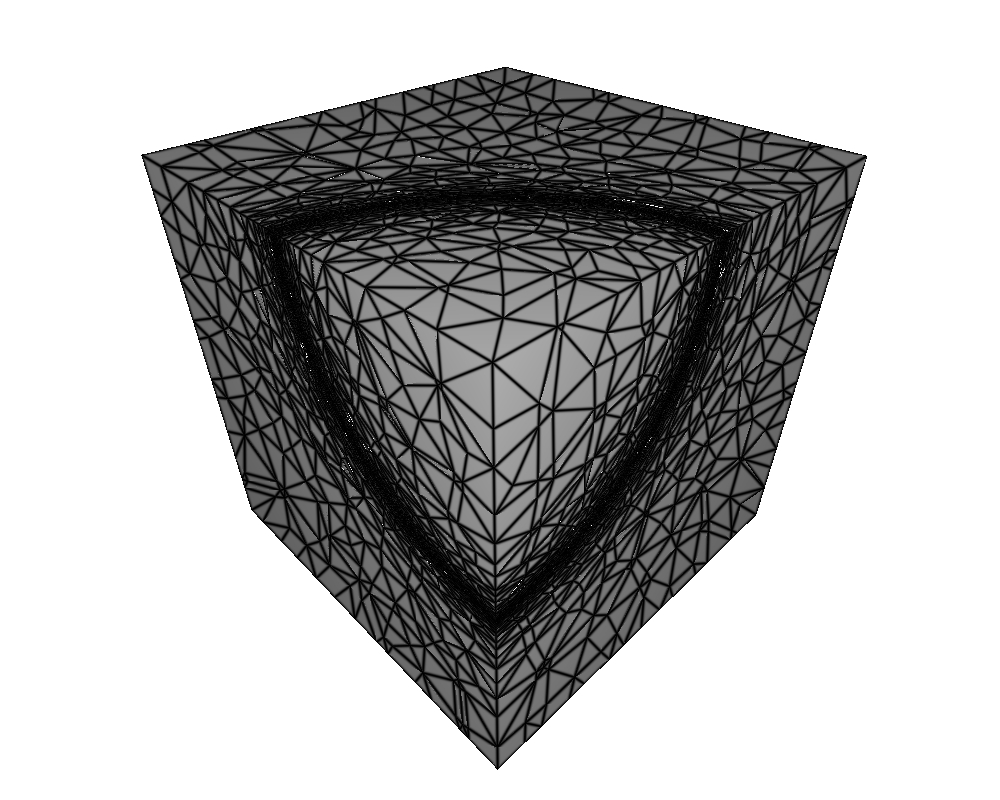}
    \caption{$t = 0.25$}
  \end{subfigure}
  \begin{subfigure}{0.32\textwidth}
    \includegraphics[width=\textwidth]{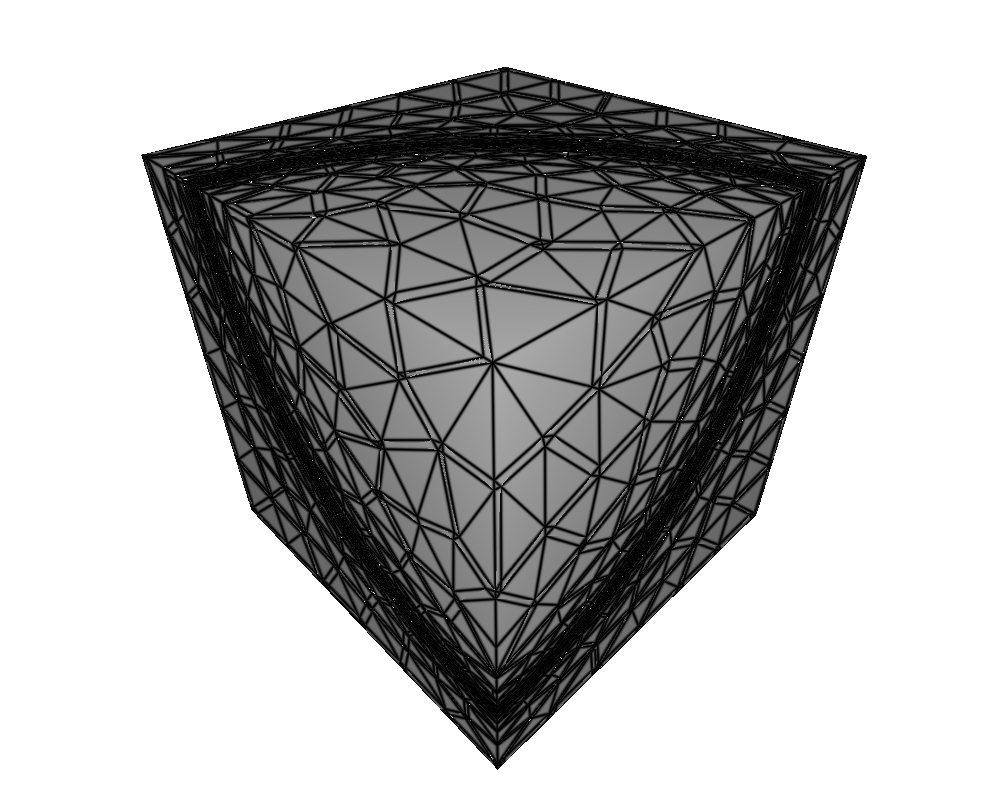}
    \caption{$t = 0.95$}
  \end{subfigure}
  \caption{Visualization of the pentatopal meshes used to evaluate the performance of the visualization system.}
  \label{fig:pentatope-vis}
\end{figure}

Sample visualizations produced by the current visualization algorithm for these meshes are shown in Fig.~\ref{fig:pentatope-vis} for $t = 0$, $t = 0.25$ and $t = 0.95$ seconds.

\begin{figure}
  \centering
  \includegraphics[width=0.8\textwidth]{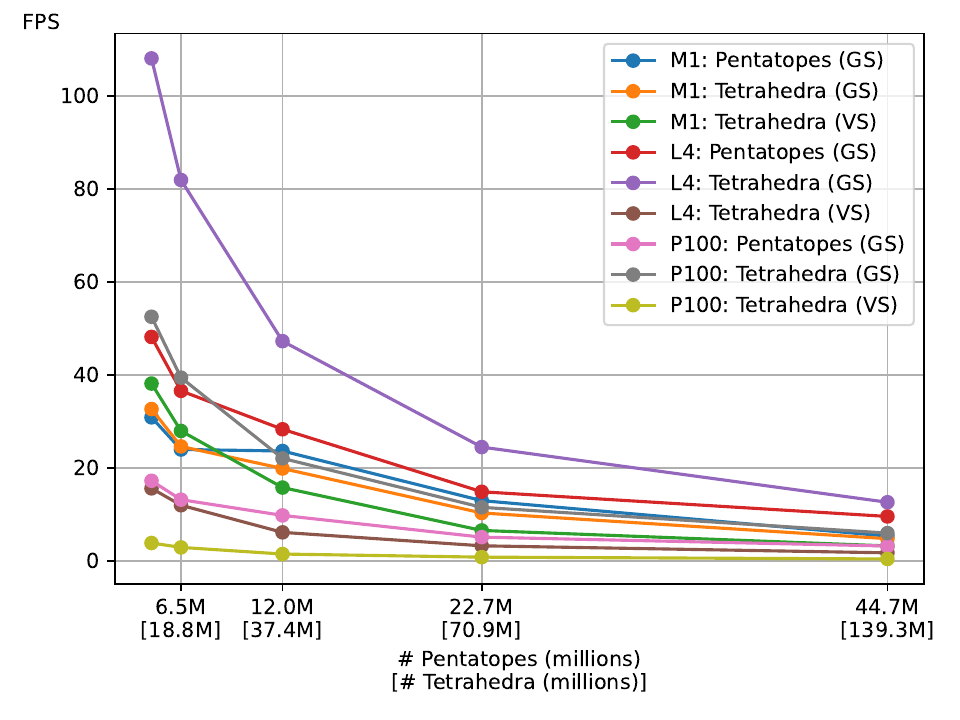}
  \caption{Frame rate obtained when rendering the pentatope meshes with the geometry-shader- and vertex-shader-based approaches on three GPUs.}
  \label{fig:fps-pentatopes}
\end{figure}

Again, the rendering time was sampled 50 times for the various approaches: (1) the geometry-shader-based approach applied directly to the pentatopes, (2) the geometry-shader-based approach applied to the unique tetrahedra extracted from the pentatope mesh and (3) the vertex-shader-based approach applied to the unique tetrahedra extracted from the original pentatope mesh.
The frames per second for each approach (with meshes of various resolutions) are plotted in Fig.~\ref{fig:fps-pentatopes} - some of these meshes were obtained as uniform subdivisions of the pentatopes of the original meshes produced by \texttt{avro}.
Note that the horizontal axis also labels the number of extracted tetrahedra of the pentatopal meshes in parantheses.
Similar to the results for the wing-flap and wind-turbine meshes, the geometry-shader-based approaches generally perform better than the vertex-shader-based approach.
On an Apple M1 Pro, the frame rate is roughly 30 frames per second for the 12M pentatope mesh (37.4M tetrahedra), and 22 frames per second when using an NVIDIA P100 GPU.
The geometry-shader-based approach with an NVIDIA L4 GPU achieves a frame rate of over 20 frames per second for the 22.7M pentatope mesh (70.9M tetrahedra).

\section{Perspectives}

This paper presented and demonstrated solutions to two problems necessary for enabling future research in $3d+t$ ($4d$) spacetime meshing.
The first solution addresses the need for closed tetrahedral meshes of moving geometries.
Each $n$-dimensional geometry entity (CAD Edge or Face) at two time step locations were mapped to the parameter space of the entity and augmented to a $(n+1)$ dimensional space where either triangulations (Edges) or tetrahedralizations (Faces) were computed.
This approach allows the resolution of the geometry to vary between time steps.
The method was demonstrated on simple geometries as well as more complex geometries with rigid-body transformations, and could also be used if the surface definitions change in time.

Next, two solutions for rendering spacetime meshes (either tetrahedral or pentatopal meshes) were presented.
These consisted of vertex-shader- and geometry-shader-based approaches for computing the mesh-hyperplane intersections directly in the rendering pipeline.
The approaches using a geometry shader performed best, achieving frame rates of about $20 - 30$ frames per second for relatively large meshes (about $50$ million tetrahedra).

Though the tessellation method was successful for the geometries studied in this work, future work should address some open problems.
First, the triangles in the parameter spaces of each entity can be poorly shaped, particularly if the geometry contains singularities or if the face is bounded by a degenerate edge.
These issues could be solved by either improving the triangulations in $(u, v)$ space, or by introducing an intermediate mapping of the entity to improve the triangle quality for the constrained tetrahedralization algorithm.
Furthermore, although the number of Steiner vertices is kept to a minimum (only introduced if necessary by the constrained tetrahedralization algorithm), the calculation of the $4d$ coordinates of these vertices could be improved.
Currently, a weighted average of the coordinates at each bounding time step is used, but this does not necessarily lie on the geometry at the precise time coordinate.
An analytic description of the geometry in between time steps is not known, but the parametrization of each surface patch could be augmented to a trivariate spline representation~\cite{Cohen_2001} to obtain a complete parametrization of the surfaces in time.

Future work could consist of extending the meshing algorithm to handle topology changes in the geometry, e.g. if a hole were created, or a body appears/disappears at a particular time.
It would also be worthwhile to develop a pentatope meshing algorithm that respects the constraints defined by the tetrahedralizations produced by the current meshing algorithm.
Though the focus here is on producing meshes for visualization, the resulting meshes are conforming and techniques for improving the quality of the tetrahedra could also be explored.

\section*{Acknowledgments}

Many thanks to Bob Haimes for reviewing a draft of this manuscript and for answering questions about \texttt{EGADS}.

\section*{Funding sources}

This research did not receive any specific grant from funding agencies in the public, commercial, or not-for-profit sectors.

\bibliography{references}
\end{document}